\documentclass[%
 aip,
 amsmath,amssymb,
 reprint,%
]{revtex4-1}

\usepackage{graphicx}% Include figure files
\usepackage{dcolumn}% Align table columns on decimal point
\usepackage{bm}% bold math
\usepackage[utf8]{inputenc}
\usepackage[T1]{fontenc}
\usepackage{mathptmx}
\usepackage{textcomp}
\usepackage{gensymb}
\usepackage[none]{hyphenat}

\begin{document}

\preprint{AIP/123-QED}

\title{Thermal and Oxidation Stability of Ti\textsubscript{x}W\textsubscript{1-x} Diffusion Barriers Investigated by Soft and Hard X-ray Photoelectron Spectroscopy}
% Force line breaks with \\

\author{C. Kalha}
\affiliation{Department of Chemistry, University College London, 20 Gordon Street, London, WC1H~0AJ, United Kingdom.}

\author{S. Bichelmaier}%
\altaffiliation[Current address: ]{Technische Universität Wien, Department of Computational Chemistry, Getreidemarkt 9/165, 1060 Vienna, Austria.}
\affiliation{Kompetenzzentrum Automobil- und Industrie-Elektronik GmbH, Europastraße 8, 9524 Villach, Austria.}

\author{N.~K.~Fernando}
\affiliation{Department of Chemistry, University College London, 20 Gordon Street, London, WC1H~0AJ, United Kingdom.}%Lines break automatically or can be forced with \\

\author{J.~V.~Berens}
\affiliation{Kompetenzzentrum Automobil- und Industrie-Elektronik GmbH, Europastraße 8, 9524 Villach, Austria.}

\author{P.~K.~Thakur}%
\author{T.~-L.~Lee}%
\affiliation{Diamond Light Source Ltd., Harwell Science and Innovation Campus, Didcot, Oxfordshire, OX1 3QR, United Kingdom.}

\author{J.~J.~Gutiérrez Moreno}
\affiliation{Barcelona Supercomputing Center (BSC).}

\author{S.~Mohr}
\affiliation{Barcelona Supercomputing Center (BSC).}
\affiliation{Nextmol (Bytelab Solutions SL), C/ Roc Boronat 117, 08018 Barcelona, Spain.}

\author{L.~E.~Ratcliff}
\affiliation{Department of Materials, Imperial College London, London SW7 2AZ, United Kingdom.}

\author{M.~Reisinger}
\author{J.~Zechner}
\author{M.~Nelhiebel}
\affiliation{Kompetenzzentrum Automobil- und Industrie-Elektronik GmbH, Europastraße 8, 9524 Villach, Austria.}

\author{A.~Regoutz}
 \email{a.regoutz@ucl.ac.uk}
\affiliation{Department of Chemistry, University College London, 20 Gordon Street, London, WC1H~0AJ, United Kingdom.}%Lines break automatically or can be forced with \\

\date{\today}% It is always \today, today,

\begin{abstract}
The binary alloy of titanium-tungsten (TiW) is an established diffusion barrier in high-power semiconductor devices, owing to its ability to suppress the diffusion of copper from the metallisation scheme into the surrounding silicon substructure. However, little is known about the response of TiW to high temperature events or its behaviour when exposed to air. Here, a combined soft and hard X-ray photoelectron spectroscopy (XPS) characterisation approach is used to study the influence of post-deposition annealing and titanium concentration on the oxidation behaviour of a 300~nm-thick TiW film. The combination of both XPS techniques allows for the assessment of the chemical state and elemental composition across the surface and bulk of the TiW layer. The findings show that in response to high-temperature annealing, titanium segregates out of the mixed metal system and upwardly migrates, accumulating at the TiW/air interface. Titanium shows remarkably rapid diffusion under relatively short annealing timescales and the extent of titanium surface enrichment is increased through longer annealing periods or by increasing the precursor titanium concentration. Surface titanium enrichment enhances the extent of oxidation both at the surface and in the bulk of the alloy due to the strong gettering ability of titanium. Quantification of the soft X-ray photoelectron spectra highlights the formation of three tungsten oxidation environments, attributed to WO\textsubscript{2}, WO\textsubscript{3} and a WO\textsubscript{3} oxide coordinated with a titanium environment. This combinatorial characterisation approach provides valuable insights into the thermal and oxidation stability of TiW alloys from two depth perspectives, aiding the development of future device technologies.
\end{abstract}

\maketitle

%%%%%%%%%%%%%%%%%%%%%%%%%%%%%%%%%%%%%%%%%%%%%%%%%%%%%%%%%%%%%%%%%%%%%%%%%%%
\section{Introduction}
%%%%%%%%%%%%%%%%%%%%%%%%%%%%%%%%%%%%%%%%%%%%%%%%%%%%%%%%%%%%%%%%%%%%%%%%%%%

The transition from aluminium to copper metallisation schemes in semiconductor technologies was one of the major breakthroughs for the industry, as it offered significant improvements in performance while also enabling a reduction in feature sizes.~\cite{IBM} However, a major challenge is the fast diffusion of copper toward the silicon device at relatively low temperatures (<200~$\degree$C), which can lead to the development of electrically active deep traps,~\cite{Sachdeva_2001} and even the formation of copper silicide phases,~\cite{Cros1990, STOLT1991147} both of which can promote device failure. To counter this negative interaction, device architectures have been modified to incorporate a thin metallic layer between the metallisation and silicon substructure, known as a diffusion barrier. A binary pseudo-alloy of titanium and tungsten (TiW) has long been a viable diffusion barrier candidate for semiconductor devices due to its strong blocking efficiency toward metallisation schemes, high thermal breakdown strength, and good chemical inertness.~\cite{Cunningham_1970, NICOLET1978415} A significant proportion of research into TiW diffusion barriers was driven by the introduction of copper metallisation lines in very-large-scale and ultra-large-scale integrated circuits.~\cite{GHATE1978117, OPAROWSKI1987313, DIRKS1990201, Alay1991, Wang_SQ_1993, Chiou_1995, Rosenberg2000, Chang_2000, Kaloyeros}
The particular focus of these investigations involved determining the barrier failure temperature through rapid thermal annealing (RTA) treatments, followed by characterisation to obtain elemental depth profiles and detect silicide phases to assess the failure modes. However, these tests often prove inconclusive or are not transferable as the short annealing timescales in RTA treatments do not reflect the duration of high temperature exposure seen during manufacture or service. Research into TiW has since seen a resurgence in response to these limitations, but also due to the development of wide band-gap materials for power semiconductor devices and the increasingly higher temperatures experienced in high power applications.~\cite{ROSHANGHIAS2014386, FUGGER20142487, Plappert_2012, Berens_2020} Despite the effectiveness of TiW as a copper diffusion barrier, when subjected to high temperatures for a prolonged duration, whether experienced during the many thermal treatments in the fabrication route or during operation, the barrier is susceptible to degradation. During such events, titanium exhibits the ability to segregate out of the barrier layer, diffusing into the metallisation.~\cite{Baeri_2004, FUGGER20142487} The addition of titanium to tungsten improves the diffusion barrier performance, increasing its adhesion to the surrounding layers,~\cite{TING1982327, Furuya_1998} enhancing the corrosion resistance through the formation of stable oxides,~\cite{Cunningham_1970} and increasing the barrier failure temperature relative to a single tungsten barrier.~\cite{Christou_1975,Lane_1991,Wang_SQ_1993} Therefore, the depletion of titanium out of the barrier could be detrimental, resulting in an increased susceptibility of barrier degradation, and consequently allowing copper to interact with the silicon substructure.\par

Despite the extensive research into this material to date, expansion of the research into other equally important areas, such as this detrimental titanium diffusion mechanism or the oxidation behaviour of TiW alloys is limited or not investigated from a device perspective.\cite{FERRONI1997499, SIOL202095} The oxidation behaviour of TiW alloys is of particular interest in this study as it will have a synergistic relationship with the already observed titanium diffusion process. The interaction of oxygen with these TiW alloys can occur similarly either during the manufacturing stages or service of the device. Several studies have explored the forced oxidation of the TiW barrier prior to copper deposition by exposing the layer to the ambient environment (e.g.\ air) via a vacuum break. The exposure to air is assumed to decorate the TiW grain boundaries with oxides, retarding the diffusion of copper through defect sites within the barrier.\cite{BAKER198053, Olowolafe1985, Palmstron} Additionally, and relevant to industrial applications, these devices consist of a thick copper interconnect scheme deposited on top of the TiW barrier layer. The process used to deposit such copper interconnects often relies on organic/inorganic precursors and additives, \cite{West2003} all of which can result in the inclusion of oxygen impurities into the copper.\cite{Lee_2018} During service, the residual oxygen within the material may diffuse downward toward the TiW/Cu interface, interacting with the TiW layer. In addition, trace amounts of oxygen present in the deposition chamber prior to copper deposition are a further source of oxygen. All of these TiW oxygen exposure routes can significantly influence the performance of the device. Therefore, it is imperative to understand these TiW alloys including the development of any potential interfacial oxide that may form, as well as the behaviour of the alloy in response to high temperature exposure.\par

\begin{figure*}
\centering
    \includegraphics[keepaspectratio, width= 17 cm]{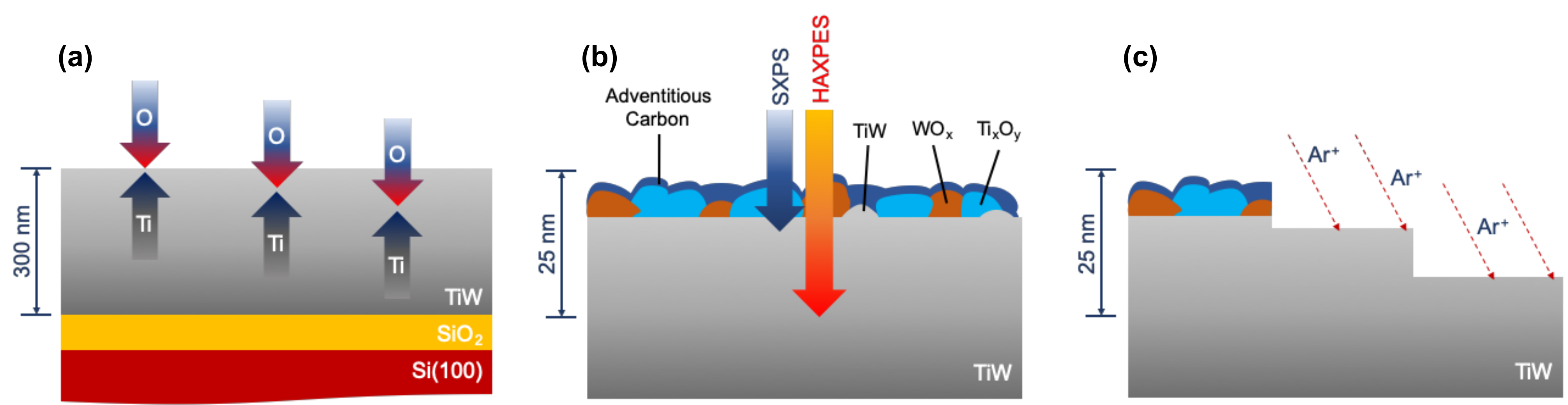}
    \caption{Schematic profiles of Si/SiO\textsubscript{2}/TiW device stacks. (a) A simplified model of the titanium and oxygen diffusion process within TiW. It is assumed that Ti will diffuse from within the bulk and accumulate at the free surface, whereas oxygen is assumed to penetrate the TiW both at the surface and within the bulk. (b) A detailed TiW surface, showcasing the assumed nature of the oxidation scale and adventitious carbon surface layers. The blue and orange arrows in the centre illustrate the probing depth of SXPS compared to HAXPES, respectively. (c) An illustration of the SXPS depth profile, (L to R) the as-received sample, a first sputtering cycle and then a second sputtering cycle.}
    \label{fig:Diagram}
\end{figure*}

To gain a more robust insight into the nature of this barrier, the effects of post-deposition annealing and titanium concentration on the electronic structure and chemical state of TiW thin films deposited on a silicon wafer substrate are explored. Moreover, by exposing these films to the ambient environment, the titanium diffusion mechanism in parallel to the oxidation behaviour is studied (schematically summarised in Fig.~\ref{fig:Diagram}(a)). Combining laboratory-based soft X-ray photoelectron spectroscopy (SXPS) and synchrotron-based hard X-ray photoelectron spectroscopy (HAXPES) a systematic study across the surface and bulk of the TiW layer is conducted, \cite{Woicik2016} allowing for the assessment of this material at multiple depths, as illustrated in Fig.~\ref{fig:Diagram}(b). Additional support is also provided by density functional theory (DFT)~\cite{Hohenberg1964,Kohn1965} to aid discussion of the influence of the oxidation process on the electronic structure of the alloys.\par

%%%%%%%%%%%%%%%%%%%%%%%%%%%%%%%%%%%%%%%%%%%%%%%%%%%%%%%%%%%%%%%%%%%%%%%%%%%
\section{\label{sec:Method}Methodology}
%%%%%%%%%%%%%%%%%%%%%%%%%%%%%%%%%%%%%%%%%%%%%%%%%%%%%%%%%%%%%%%%%%%%%%%%%%%

%%%%%%%%%%%%%%%%%%%%%%%%%%%%%%%%%%%%%%%%%%%%%%%%%%%%%%%%%%%%%%%%%%%%%%%%%%%
\subsection{\label{sec:EXP_method}Experimental Methods}
%%%%%%%%%%%%%%%%%%%%%%%%%%%%%%%%%%%%%%%%%%%%%%%%%%%%%%%%%%%%%%%%%%%%%%%%%%%
Si(100)/SiO\textsubscript{2}(50~nm)/TiW(300~nm) were prepared using standard industrial processes with the TiW layer fabricated by magnetron sputter deposition. By varying the sputter parameters four different titanium concentrations of $\approx$ 11, 17, 22 and 27~at.\% could be realised in the coatings. The concentrations are regularly monitored using X-ray fluorescence (XRF). XRF is a bulk-sensitive chemical characterisation techniques, therefore, these concentrations are taken as bulk concentrations. Each of the four samples was then annealed at 400 $\degree$C under an inert forming gas atmosphere (5\% H\textsubscript{2}, 95\% N\textsubscript{2}) for two different timescales - 0.5~h and 5~h. The annealing duration does not take into account the time taken for the annealing temperature to ramp up or cool down. An as-deposited (i.e.\ no anneal) sample (AD) was used as a reference. After annealing, samples were removed from the chamber and were left exposed to the ambient environment until measurements were conducted. Although residual oxygen may be present in the deposition chamber, it is assumed that oxidation within the deposition chamber is negligible.\par

SXPS measurements were conducted using a laboratory-based Thermo Fisher Scientific K-Alpha spectrometer, which features a monochromatic Al~K$\alpha$ (1486.7~eV) X-ray source, a 180$\degree$ double focusing hemispherical analyser and a two-dimensional detector. The X-ray source operated under a 12~keV anode bias and 6~mA emission current and a 400~$\mu$m elliptical spot size was selected for all measurements. Pass energies of 15, 20 and 200~eV were used for core level, valence band, and survey spectra acquisition, respectively, and the base pressure of the system was 2$\times$10\textsuperscript{-9}~mbar. A flood gun was utilised to prevent sample charging. For depth profiles, samples were sputtered using a focused 0.5~keV Ar\textsuperscript{+} ion source. Two sputter cycles were completed, each lasting 300~s (5~minutes). The sputter area was approximately 2~x~2~mm\textsuperscript{2} and subsequent measurements were taken at the centre of the sputter crater. This depth profile is schematically illustrated in Fig.~\ref{fig:Diagram}(c).\par

HAXPES measurements were conducted at beamline I09 at Diamond Light Source, UK.\cite{Beam2018} An X-ray excitation energy of 5922~eV (further referred to as 5.9~keV for simplicity) was selected using a Si~(111) double crystal monochromator and an additional Si~(004) post channel-cut monochromator. The beamline operates with a VG Scienta EW4000 electron energy analyser that has a $\pm$28$\degree$ acceptance angle. Samples were measured using grazing incidence (4$\degree$) and near normal emission geometry to maximise the signal intensity. A pass energy of 100~eV was used for core level and survey scan acquisition, whereas a 200~eV pass energy was used to collect high resolution valence band spectra. The energy resolution of soft and hard XPS instruments is approximately 300 meV and 250 meV, respectively, determined by measuring the Fermi edge of a polycrystalline gold foil.\par

For both SXPS and HAXPES measurements, survey, key core level (W~4\textit{f}, O~1\textit{s}, Ti~2\textit{p}, and C~1\textit{s}), and valence band spectra were acquired. All spectra were charge corrected by aligning the binding energy scale to the intrinsic Fermi edge of their respective samples. All core level spectra were normalised to the maximum intensity of the W~4\textit{f}\textsubscript{7/2} peak. Valence band spectra were normalised to the total valence band area after the removal of a Shirley-type background from the spectra. All peak fit analysis was carried out using the Thermo Avantage\textsuperscript{TM} software package.\par

\begin{figure}[ht!]
\centering
    \includegraphics[keepaspectratio, width= 8.5 cm]{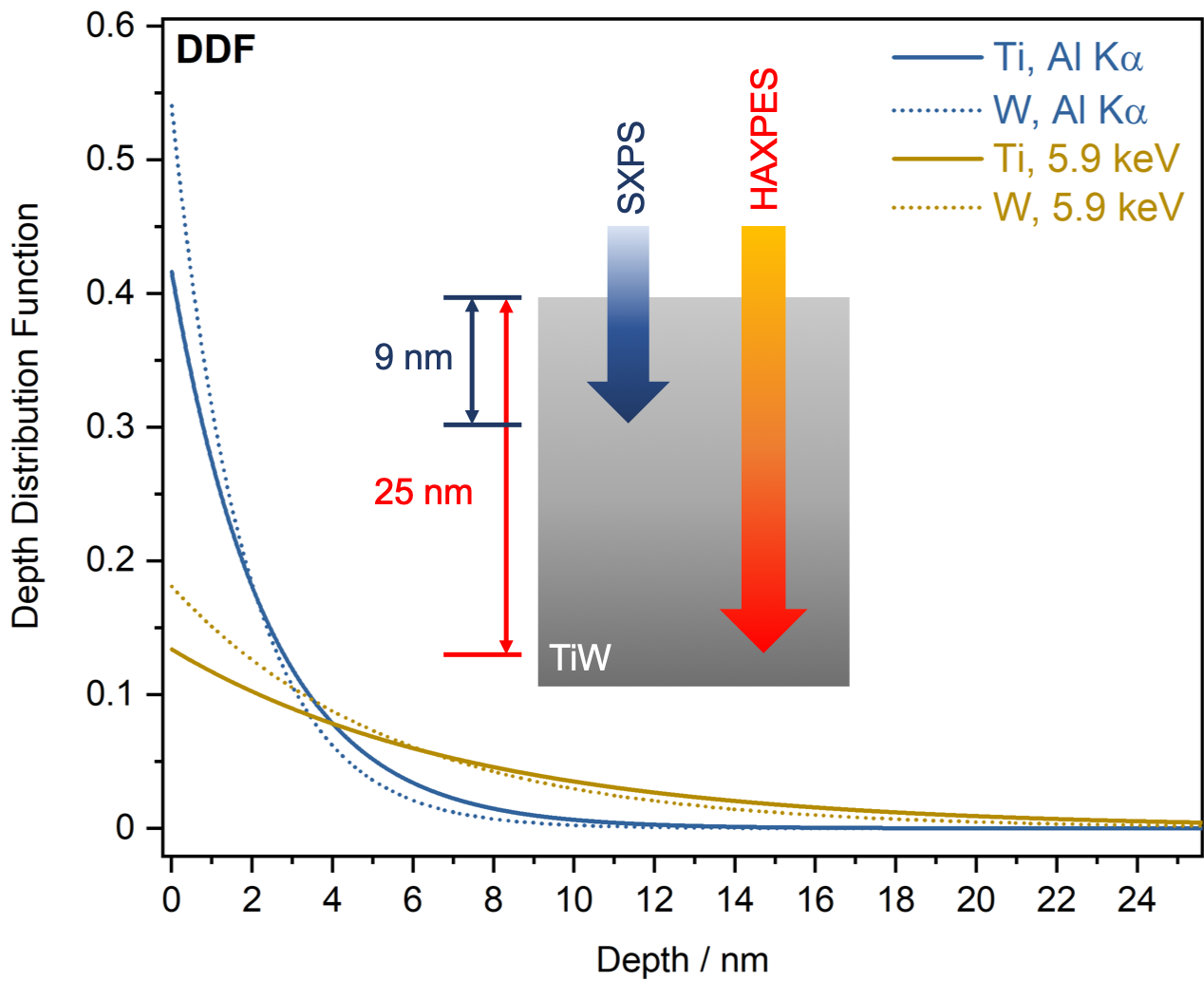}
    \caption{DDF for Ti and W metal, calculated for the two photon energies used (Al K$\alpha$ and 5.9 keV), with the inset illustrating the differences in the probing depth for SXPS versus HAXPES.}
    \label{fig:DDF}
\end{figure}

SXPS provides surface-sensitive information, whereas HAXPES is able to access the "bulk" and hence probe deeper into the sample, without the need of sample preparation. \cite{Woicik2016} A depth distribution function (DDF) is a useful measure that allows for the determination of the probing depth of each technique giving an indication of the origin of the photoelectron signal intensity. The DDF is a derivation of the Beer-Lambert Law, described in detail by Berens \textit{et al}.\ \cite{Berens_2020} and by integrating the DDF over a particular depth range, the percentage of the total signal from these elements can be determined. The DDF was calculated for titanium and tungsten metal using the Al K$\alpha$ and 5.9~keV photon energies and the maximum inelastic mean free path (IMFP) values reported by Shinotsuka \textit{et al}.,\cite{Shinotsuka2015} with the results displayed in Fig.~\ref{fig:DDF}. The DDF relies heavily on the IMFP and hence is a function of the material (e.g.\ density, number of valence electrons, atomic mass etc.), which is why titanium and tungsten show minor differences in the distribution. Through integrating the DDF function across the top 5~nm for tungsten metal, the clear difference in information depth becomes apparent. Whilst the majority of the signal in SXPS, approximately 93.1\% of the total signal, originates from within this window, only 59.5\% of signal is accessed in HAXPES with the remainder stemming from deeper in the sample. As mentioned above, due to the elemental differences in DDF, titanium shows values of 87.4\% and 48.4\% signal contribution from the top 5~nm in SXPS and HAXPES, respectively. In the case of the TiW alloys investigated in this line of work, the titanium concentration is low ($<$30~at.\%) and therefore, using the tungsten DDF is a good estimate for these samples. The DDF overall highlights the extreme surface-sensitivity of SXPS, with the majority of the signal intensity originating from the top few nanometres of the surface, compared to the surface-insensitivity of HAXPES, in which the majority of the signal originates from within the bulk.\par

%%%%%%%%%%%%%%%%%%%%%%%%%%%%%%%%%%%%%%%%%%%%%%%%%%%%%%%%%%%%%%%%%%%%%%%%%%%
\subsection{\label{sec:Theory_method}Theoretical Methods}
%%%%%%%%%%%%%%%%%%%%%%%%%%%%%%%%%%%%%%%%%%%%%%%%%%%%%%%%%%%%%%%%%%%%%%%%%%%

The electronic structure of the occupied states of a material can be explored through the collection of valence band spectra. However, to arrive at a complete description complementary theoretical calculations are essential. Here, DFT with a linear scaling (LS) approach is employed as implemented in the wavelet-based BigDFT code to calculate the electronic structure of metallic tungsten and compare with these tungsten dominated alloys.~\cite{Genovese_2008, Mohr2014, Mohr2015, Ratcliff2020} Although bulk tungsten does not have a large enough unit cell to require a LS treatment, LS-BigDFT has previously been demonstrated to give excellent agreement with more traditional cubic scaling approaches for tungsten.~\cite{Mohr2018} Moreover, the localised atom-centred support function basis of LS-BigDFT provides an intuitive approach for generating the partial density of states (PDOS) of tungsten metal.\par

A single-point calculation was performed for a $9\times 9\times 9$ (1458 atom) supercell at the $\Gamma$-point only. A perfect body centered cubic (BCC) structure with 3-dimensional periodic boundary conditions and a DFT equilibrium lattice parameter of 3.23~Å was adopted. Calculations employed a wavelet grid spacing of 0.38~bohr, coarse and fine radius multipliers of 6 and 8, respectively, and 9 support functions per tungsten atom with localisation radii of 7.5~bohr. The cutoff for the density kernel was set to 11.0~bohr. The Krack HGH pseudopotential with the W~6\textit{s}\textsuperscript{2} and 5\textit{d}\textsuperscript{4} valence electrons,~\cite{krack2005pseudopotentials} and the PBE exchange correlation functional were used.~\cite{Perdew1996} The Fermi Operator Expansion (FOE) method,~\cite{FOEPhysRevLett.73.122,FOEPhysRevB.51.9455} implemented via the CheSS library,~\cite{Mohr2017a} was used to optimize the density kernel. A finite electronic temperature of 0.136~eV was introduced into the calculations to facilitate the convergence of the metallic system. The PDOS was calculated using a Mulliken-type projector onto the support functions.~\cite{Mohr2017b, Dawson2020}\par

To allow for direct comparison with experimental data, the calculated PDOS of tungsten metal was broadened to match the experimental broadening due to factors such as instrumental resolution and temperature effects. The level of broadening is determined by fitting the intrinsic Fermi edge width of a measured polycrystalline gold foil using a Gaussian-Dirac step function (i.e.\ the experimental resolution). Broadening attributed to these factors is best described using a Gaussian function. Gaussian smearing of 0.30~eV and~0.25 eV was used to compare to the valence bands collected with the Al~K$\alpha$ and 5.9~keV excitation energies, respectively. Following the broadening, the PDOS was weighted by the one-electron photoionisation cross sections to account for the photon energy dependence of the experimental spectra. Photoionisation cross section weighting of the PDOS was applied using the Galore software package which extracts values from the tabulated data calculated by Scofield and divides the orbital energies with their respective cross section at the given photon energy (Al K$\alpha$ or 5.9~keV).\cite{Jackson2018, Scofield1973, Dig_Sco_2020} The total density of states (TDOS) is then calculated by summing the PDOS. The DFT theoretical spectra were normalised to the maximum intensity of the experimental valence band spectra.\par

%%%%%%%%%%%%%%%%%%%%%%%%%%%%%%%%%%%%%%%%%%%%%%%%%%%%%%%%%%%%%%%%%%%%%%%%%%%
\section{\label{sec:Results}Results and Discussion}
%%%%%%%%%%%%%%%%%%%%%%%%%%%%%%%%%%%%%%%%%%%%%%%%%%%%%%%%%%%%%%%%%%%%%%%%%%%

The survey spectra collected using SXPS and HAXPES (see Supplementary Information) show the presence of titanium, tungsten, oxygen and carbon for all samples. SXPS measurements detect carbon with a significantly higher intensity compared to HAXPES, expected from the exposure of the samples to the ambient environment and the formation of a surface carbon layer, known as adventitious carbon, as illustrated in Fig.~\ref{fig:Diagram}(b). HAXPES, as mentioned above, is bulk-sensitive and therefore this surface carbon layer does not contribute significantly to the total signal. SXPS also detects minor traces of nitrogen, sodium and calcium in a small number of samples. Sodium and calcium are assumed to be from external contamination during sample preparation and handling. Nitrogen is assumed to be from an organic carbon-based impurity due to the exposure to the ambient environment.\par

\subsection{\label{sec:Annealing_D}Influence of Post-Deposition Annealing}
%%%%%%%%%%%%%%%%%%%%%%%%%%%%%%%%%%%%%%%%%%%%%%%%%%%%%%%%%%%%%%%%%%%%%%%%%%%

\begin{figure*}
\centering
    \includegraphics[keepaspectratio, width= 15 cm]{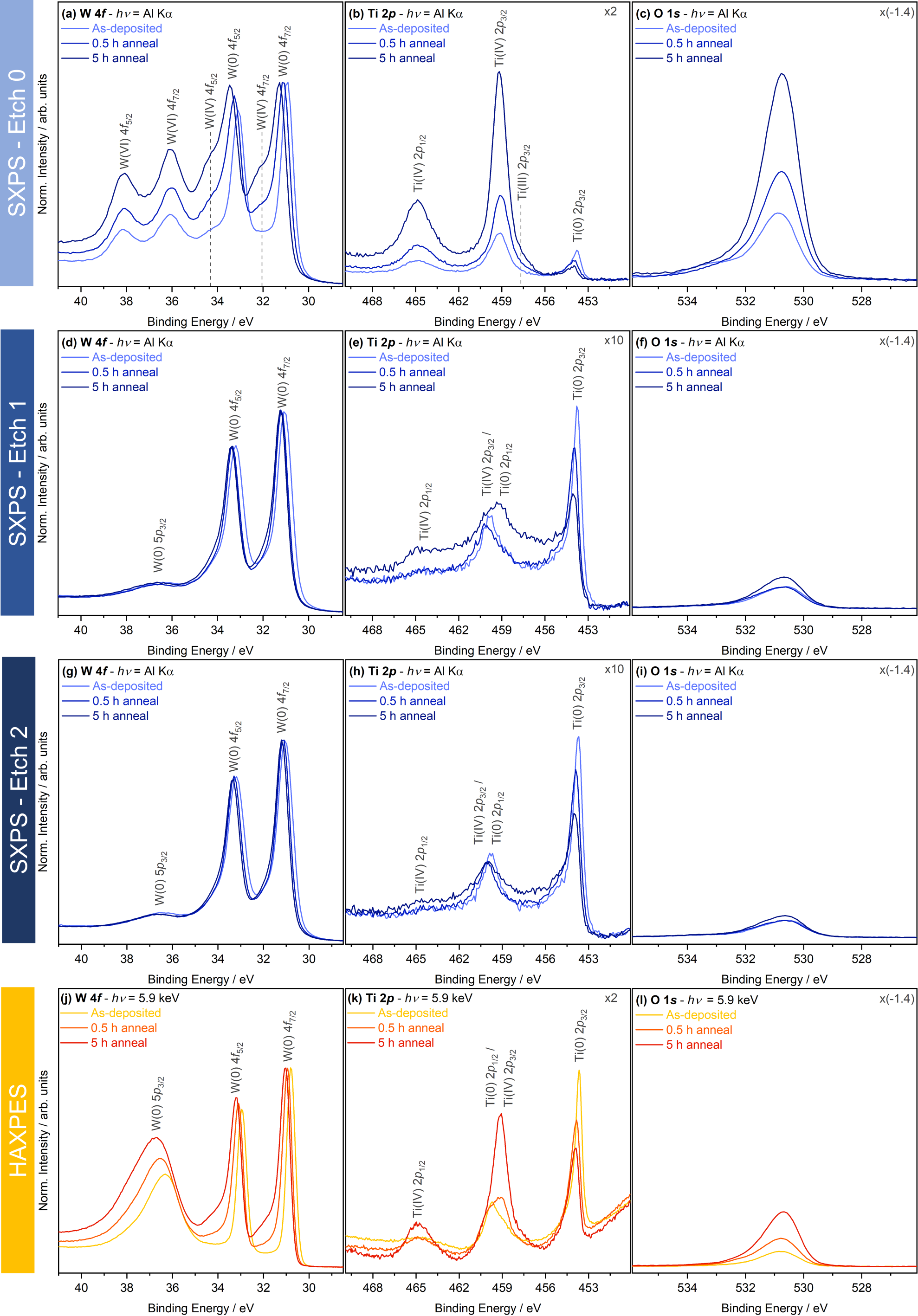}
    \caption{Key core level spectra acquired using SXPS ($h\nu$~= Al K$\alpha$) and HAXPES ($h\nu$~= 5.9~keV) for a sample containing 27 at.\% Ti as a function of annealing duration. From left (L) to right (R) - W~4\textit{f}, Ti~2\textit{p}, and O~1\textit{s}. From the top row to the bottom - initial SXP spectra, spectra acquired after sputter removal for 5~minutes, spectra acquired after sputter removal for 10~minutes, and finally, spectra acquired on as-received samples using HAXPES. All spectra are normalised to the maximum intensity of the W 4\textit{f} spectra. The magnification of the spectra are presented relative to the W 4\textit{f} y-axis scale.}
    \label{fig:Annealing}
\end{figure*}

Fig.~\ref{fig:Annealing} displays the W~4\textit{f}, Ti~2\textit{p}, and O~1\textit{s} core level spectra acquired with SXPS, SXPS depth profiling, and HAXPES, for the sample set containing 27~at.\% Ti, highlighting the relationship between oxidation and annealing duration. SXPS measurements prior to sputtering (Etch 0) show that for both metals, significant changes in intensity are observed with increasing annealing duration. Focusing on the W~4\textit{f} spectrum, shown in Fig.~\ref{fig:Annealing}(a), four resolved peaks are present with the peaks at approximately 31~eV and 33~eV attributed to metallic tungsten (W(0)), and peaks at approximately 36~eV and 38~eV attributed to hexavalent species (WO\textsubscript{3}, W(VI)). The full list of binding energy (BE) positions can be found in the Supplementary Information and are in good agreement with literature values. \cite{Barr1978} Aside from the main peaks, a noticeable increase in intensity is observed with increasing annealing duration at +1.3 eV above the metallic W~4\textit{f} peaks. This increase in the shoulder intensity is also coupled with an increase in intensity of the hexavalent tungsten species. Additionally, the metallic W~4\textit{f} peaks exhibit a positive shift in BE of 0.1~eV after 0.5~h and 0.3~eV after 5~h of annealing. These spectral changes are also shared by the SXPS depth profile and HAXPES spectra, shown in Fig.~\ref{fig:Annealing}(d, g) and Fig.~\ref{fig:Annealing}(j), respectively. These changes are evidence that longer annealing results in an increased oxidation of the tungsten portion of the TiW alloys, moving the system from primarily metallic toward a more oxide rich environment.\par

The main contribution toward the metallic shoulder feature is assumed to be the development of tetravalent tungsten species (WO\textsubscript{2}, W(IV)). The assignment of WO\textsubscript{2} is often misinterpreted as the peak position is typically derived by linear interpolation between the metallic and hexavalent states, which predicts an approximate +3.2~eV shift relative to the metallic peak. To avoid misinterpretation, peak identification of the W~4\textit{f} core level is typically mapped to the already well-established peak assignments of the Mo~3\textit{d} core level, as both molybdenum and tungsten oxides share similar structures.\cite{GOODENOUGH1971145} For example, the Mo~3\textit{d} peaks of MoO\textsubscript{2} have been reported at a positive shift of 1.3~eV relative to the metal peaks, which has formed the basis of our assignment of the shoulders to WO\textsubscript{2}.\cite{WARREN1996345,BHATTARAI1998155,Olsson2002,Colton1976} The oxidation of these samples took place in the ambient environment; therefore, WO\textsubscript{2} and WO\textsubscript{3} are assumed to be the only tungsten oxides that exist as these are thermodynamically stable at room temperature. The formation of the lower oxidation state rather than the complete oxidation to WO\textsubscript{3} is a consequence of competitive oxidation with titanium, limiting the complete conversion to WO\textsubscript{3}. Alternatively, Warren \textit{et al}.\ reported that water acts as an oxidising agent, first converting tungsten to WO\textsubscript{2}, which then progresses to WO\textsubscript{3}, before forming an oxide hydrate. \cite{WARREN1996345} As the samples were stored in an ambient laboratory environment, the relative humidity would not have exceeded 50\% and hence would be sufficient to form WO\textsubscript{2} but insufficient to warrant a full conversion to WO\textsubscript{3} or an oxide hydrate.\cite{WARREN1996345}\par

It is clear that increasing the annealing duration results in an increase in the oxidation of tungsten. This link is attributed to the expected underlying titanium diffusion mechanism that occurs during annealing. As shown in Fig.~\ref{fig:Annealing}(b), three identifiable peaks are present in the Ti 2\textit{p} spectra collected at the surface. The metallic Ti 2\textit{p}\textsubscript{3/2} peak is located at approximately 454~eV, with the metallic Ti 2\textit{p}\textsubscript{1/2} peak hidden underneath the intense peak at 459~eV, which together with the peak at approximately 465~eV, correspond to the Ti 2\textit{p} core level of TiO\textsubscript{2} (Ti(IV)). The recorded peak positions of both Ti metal and TiO\textsubscript{2} match closely with literature values.\cite{SAYERS1978301, F19878300351, POUILLEAU1997235} The Ti(IV) peak also displays a degree of asymmetry on the lower BE side, which is more visually prominent in the 5~h annealed sample spectra, and alludes to the presence of Ti\textsubscript{2}O\textsubscript{3} (Ti(III)). With increasing annealing duration, the metallic Ti 2\textit{p} contribution is shown to diminish, while the oxide contribution is enlarged. The increase in the total spectral area of the Ti 2\textit{p} core level observed in Fig.~\ref{fig:Annealing}(b), confirms a surface enrichment of titanium driven by post-deposition annealing.\par
Annealing appears to increase the mobility of titanium atoms, leading to a thermally driven diffusion process in which titanium migrates from the bulk to the free surface, likely via defect sites (e.g.~ grain boundaries). Several studies have assessed interdiffusion phenomena experienced during high temperature annealing for systems containing titanium or titanium-based alloys.\cite{Baeri_2004, MARTINEZ20102585, Sylwestrowicz1979, EHRLICH1997122, Kai_2005}. These studies highlight that the diffusion mechanism of titanium through a polycrystalline metal over-layer proceeds in a multi-step process, initiated with diffusion from the bulk, followed by intense grain boundary diffusion and then finally accumulation at the free-surface. This process is summarised as following a type B kinetic regime (i.e.\ a combination of bulk and grain boundary diffusion).\cite{Harrison1961} Moreover, Baeri \textit{et al}. suggest that residual oxygen in the deposition chamber acts as the driving force for titanium to diffuse upward to the TiW/air interface, owing to the very strong chemical affinity of titanium.\cite{Baeri_2004} A similar finding was also shared by H{\"u}bner \textit{et al}. who observed the same process for tantalum diffusion through a copper capping layer.\cite{HUBNER2004237}
Titanium is known for its excellent gettering properties\cite{Stout1955} and is an incredibly effective oxygen scavenger, which is why it finds applications in sublimation pumps.\cite{GUPTA1975362} Therefore, if the TiW surface is enriched with titanium, greater oxidation will be expected in both metals, and this is reflected in the core level spectra when moving from the as-deposited to 5~h annealed sample. Due to the extreme oxidation that is observed, applying depth profiling to map the oxidation penetration depth brings challenges. Depth profiling was conducted via argon ion sputtering and this method will preferentially reduce any metal oxides present into sub-oxides and then into the base metal. However, the spectra shown in Fig.~\ref{fig:Annealing}(e) and Fig.~\ref{fig:Annealing}(h), which display the Ti~2\textit{p} core level after sputtering, reinforce the idea of titanium diffusion. With increasing annealing duration, the main peak intensity, which belongs to the metal contribution, decreases. This decrease indicates that under the high temperature conditions, titanium below the surface is being depleted and migrates to the surface. Of course, in these spectra, oxide contributions are shown to be minimal or mostly removed, which is a consequence of the sputtering process and reduction to the metal form. The same observation is also shared for the W~4\textit{f} and O~1\textit{s}, in that the oxide contributions are almost entirely removed. In the case of the W~4\textit{f} core level the WO\textsubscript{3} peaks are completely eliminated, revealing the previously hidden metallic W~5\textit{p}\textsubscript{3/2} peak. Subtle shifts to a positive BE for the metallic Ti~2\textit{p} and W~4\textit{f} coupled with a slight increase in intensity of the peak shoulders are observed with increasing annealing duration, reflecting both changes in the Ti:W concentration but also the influence of remnant oxide environments still present after sputtering.\par

The oxidation changes exhibited by the Ti~2\textit{p} and W~4\textit{f} surface spectra are also shared by the O~1\textit{s} spectra, with Fig.~\ref{fig:Annealing}(c) showing that with increasing annealing duration, the intensity of the O~1\textit{s} peak increases at a similar rate to the Ti~2\textit{p} intensity changes. The asymmetry of the O~1\textit{s} spectra toward higher BE indicates the presence of multiple oxygen environments.~\cite{SLEIGH199641} The main signal intensity is positioned at approximately 530 - 531~eV, which is a typical BE position of a lattice metal oxide peak.\cite{A908800H} This peak will mostly account for oxygen in titanium oxide species due to the greater level of oxidation observed relative to tungsten. This also holds true for the O~1\textit{s} spectra collected during depth profiling and with HAXPES measurements. The tail on the higher BE side is attributed to organic carbon-oxygen environments and metal-hydroxide or metal-carbonate species.\par

HAXPES measurements displayed in Fig.~\ref{fig:Annealing}(j-l), are able to reveal the true nature of the bulk TiW without the need for sample preparation via in-situ sputtering. The W~4\textit{f} spectra appear mostly metallic with a subtle shoulder present on the higher BE side of the intense W~4\textit{f} peaks, increasing in intensity with annealing duration, although the intensity is not as significant as in the case of the surface sensitive SXPS. Additionally, with increasing annealing duration, an increase in intensity of the regions between the doublet peak is observed. Both observations confirm that internal oxidation into the bulk of the TiW film occurs and that the extent of oxidation is linked to the titanium diffusion process.\par
With increasing photon energy come changes to photoionisation cross sections and this causes the metallic W~5\textit{p}\textsubscript{3/2} peak to increase in intensity relative to the W~4\textit{f} peaks. However, the peak appears slightly asymmetric and broadens with increasing annealing duration. This can be attributed to the formation of WO\textsubscript{2} and the presence of a W(IV)~5\textit{p}\textsubscript{3/2} peak underneath the intense W(0)~5\textit{p}\textsubscript{3/2} peak. This further reinforces the idea that annealing is driving titanium toward the surface, which in turn draws more oxygen to deeper depths within the TiW, leading to the partial oxidation of metallic tungsten.\par

Unlike tungsten, the Ti~2\textit{p} HAXPES spectra in Fig.~\ref{fig:Annealing}(k) display a greater presence of oxides, although metallic contributions still dominate over the oxide contributions. In-line with the previous comments, the Ti~2\textit{p} spectra reveal that samples annealed for longer timescales, result in a reduction in the metallic signal and consequently, an increase in the oxide signal. However, the rate of metallic depletion does not appear to equal the rate of oxide formation. The signal intensity of the metallic Ti~2\textit{p}\textsubscript{3/2} appears to decrease almost linearly between the three samples but the oxide Ti~2\textit{p}\textsubscript{3/2} signal appears the same for the AD and 0.5~h annealed sample, before rising to almost double the signal intensity for the 5~h annealed sample. This perhaps, suggests that once a critical titanium concentration is present, the rate of oxidation accelerates and so the titanium concentration is a limiting factor in the oxidation behaviour of these alloys.\par %The O~1\textit{s} spectra again show similarities although the intensity relative to the tungsten intensity has decreased due to the inability of oxygen to attack the bulk TiW with the same devastation as the surface.\par 

\begin{figure*}
\centering
    \includegraphics[keepaspectratio, width= 15 cm]{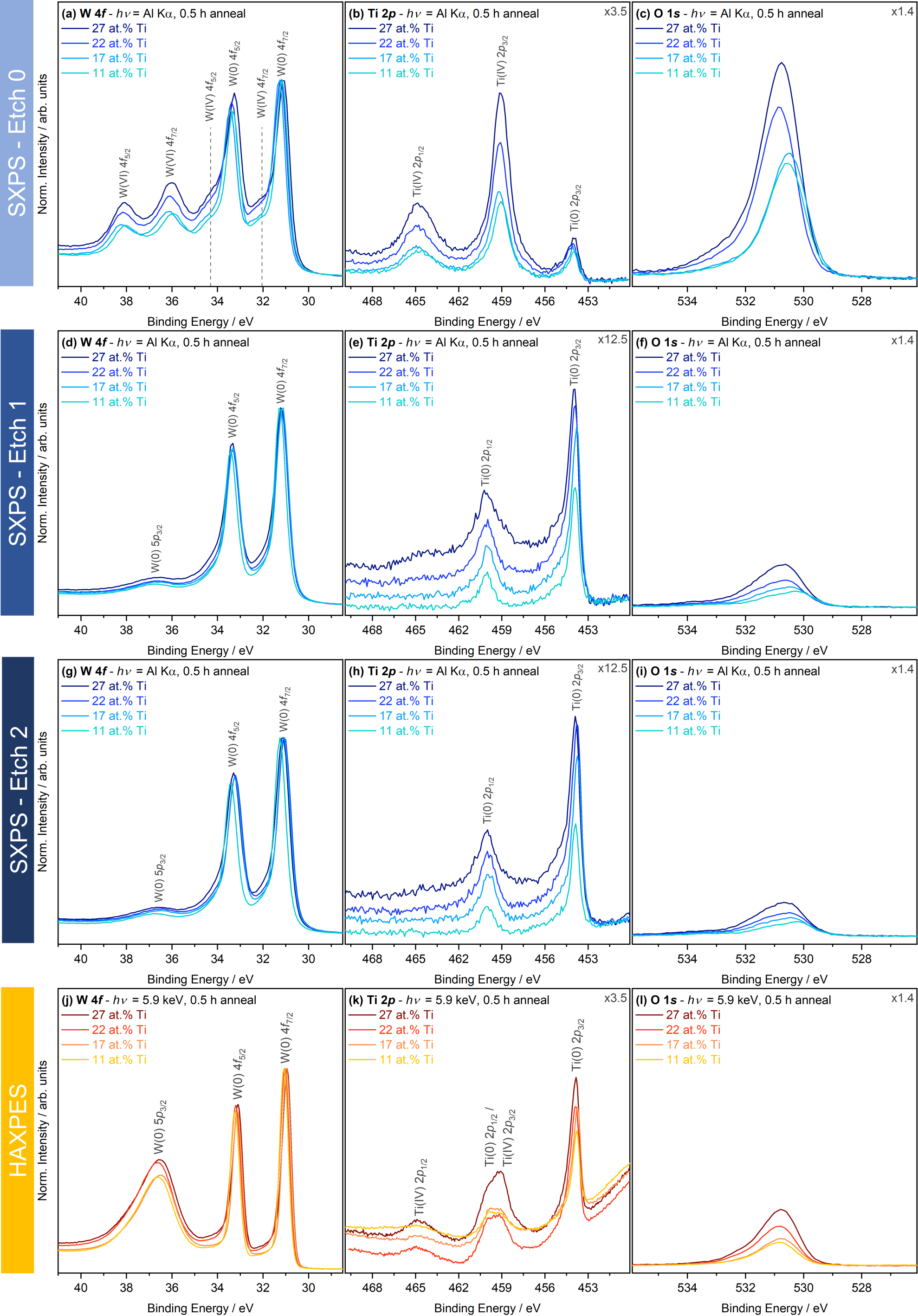}
    \caption{Key core level spectra acquired using SXPS ($h\nu$~= Al~K$\alpha$) and HAXPES ($h\nu$~= 5.9 keV) for samples with different compositions but all annealed for 0.5 h at 400$\degree$C. From left (L) to right (R) - W~4\textit{f}, Ti~2\textit{p}, and O~1\textit{s} core levels. From the top row to the bottom - initial SXP spectra, spectra acquired after sputtering for 5~minutes, spectra acquired after sputtering for 10~minutes, and finally, spectra acquired on as-received samples using HAXPES. All spectra are normalised to the maximum intensity of the W 4\textit{f} spectra. The magnification of the spectra are presented relative to the W 4\textit{f} y-axis scale.}
    \label{fig:Ti_conc}
\end{figure*}

\subsection{\label{sec:Ti_conc}Influence of Titanium Concentration}
%%%%%%%%%%%%%%%%%%%%%%%%%%%%%%%%%%%%%%%%%%%%%%%%%%%%%%%%%%%%%%%%%%%%%%%%%%%

The W~4\textit{f}, Ti~2\textit{p} and O~1\textit{s} spectra acquired with SXPS, SXPS depth profiling and HAXPES, for a range of nominal titanium concentrations (11-27~at.\%), all annealed for 0.5~h at 400~$\degree$C are depicted in Fig.~\ref{fig:Ti_conc}. As expected, oxide contributions appear to increase as the titanium concentration within the samples increases, mirroring the same phenomena that occur with increasing annealing duration. SXP spectra displayed in Fig.~\ref{fig:Ti_conc}(a-c) show that multiple oxide contributions are detected at the surface, while metallic features are mostly retained, similar to the results shown in Fig.~\ref{fig:Annealing}. This highlights that even at higher titanium concentrations, a thick oxide layer does not form and the probing depth of SXPS is still sufficient to probe through the oxide scale. Additionally, the signal intensity for metallic titanium remains fairly constant with increasing titanium concentration and it only appears that oxide contributions are significantly influenced. This finding suggests that complete oxidation is difficult, perhaps due to the competitive nature of the surface, in which two metals are competing for passivation. Alternatively, it could allude to the saturation of the surface metals with a near maximum solid solution reached, limiting further oxidation. The observed change in the Ti 2\textit{p} signal intensity when moving across the compositional range, displayed in Fig.~\ref{fig:Ti_conc}(b) appears similar to the change in intensity when increasing the annealing duration, as shown in Fig.~\ref{fig:Annealing}(b). This suggests that during annealing, a significant, almost bulk-like diffusion of titanium occurs perhaps in the range of up to 10~at.\%. Changes to the surface composition as a function of annealing duration will be discussed in greater quantitative detail in Section \ref{sec:PFA}.\par

Noticeable shifts in BE position are also reported across the three core levels, much like those reported in Fig.~\ref{fig:Annealing}. With increasing titanium concentration, the position of the Ti~2\textit{p} and O~1\textit{s} core levels shift to higher BE values, whereas the opposite effect is observed for the W~4\textit{f} core level spectra. Unlike the assumptions made in Section \ref{sec:Annealing_D}, which attribute the reported shifts to the formation of oxides, the shifts observed in Fig.~\ref{fig:Ti_conc} can be attributed to the interactions between the two base metals and perhaps the formation of intermetallic phases. Oxidation will almost always result in a positive shift in BE due to the extra Coulombic interaction between the ejected photoelectron and the ionised core hole, whereas the direction of shifts observed in intermetallic or mixed metal systems are dependent on the relative electronegativity of the constituents and the relative ratio of the constituents. The notion of an intermetallic phase formation rather than an inclusion system is difficult to confirm and is speculative without the support from other supplementary characterisation techniques. Bertoti \textit{et al}.\ reported that when going from pure nickel or pure aluminium toward binary compounds of Al-Ni and Al-Mn, noticeable shifts in BE position were observed and concluded this was indicative of the formation of an intermetallic phase.\cite{Bertoti_1992} A similar finding was also shared by García-Trenco \textit{et al}.\ in which clear systematic changes in both core level BE position and also in the valence band spectra were observed when moving from metallic palladium to a mixed palladium-indium system.\cite{GARCIATRENCO20189}\par

The O~1\textit{s} spectra across the TiW film also show similarities to Fig.~\ref{fig:Annealing} in that the main contribution is from a lattice oxide and the intensity decreases with depth. However, it is noticeable that at lower titanium concentrations, the O~1\textit{s} peak is at approximately 530.5~eV, whereas for higher concentrations a positive BE shift of 0.3~eV is observed. TiO\textsubscript{2} has reported BE values for the O~1\textit{s} core level at approximately 529.5-530~eV, \cite{A908800H,SHAM1979426,Sanjines1994, DIMITROV2002100} whereas WO\textsubscript{3} often finds itself at reported higher BE values of 530.2~-~530.8~eV. \cite{A908800H,SARMA198025,KERKHOF1978453,Colton1976, DIMITROV2002100} It has already been established that a greater titanium concentration at the surface results in greater oxidation of both metals but in particular enhances the tungsten oxidation rate, therefore, this observed shift in the O~1\textit{s} BE for higher titanium concentrations could be indicative of the higher concentration of tungsten oxide states.\par

HAXPES results showcased in Fig.~\ref{fig:Ti_conc}(j-l), similar to those discussed in Section \ref{sec:Annealing_D}, tell a different story to the SXPS depth profiling results, as they highlight that oxidation does occur in the bulk but that metallic contributions dominate. The depth profiling results (Fig.~\ref{fig:Ti_conc}(d-i)) show as expected that with increasing titanium concentration, an increase in the Ti~2\textit{p} intensity increases and this is coupled with an increase in the O~1\textit{s} spectra, as alloys with a greater titanium concentration allow for greater oxidation.\par

\subsection{\label{sec:PFA}Peak Fit Analysis and Quantification}
%%%%%%%%%%%%%%%%%%%%%%%%%%%%%%%%%%%%%%%%%%%%%%%%%%%%%%%%%%%%%%%%%%%%%%%%%%%

\begin{figure*}
\centering
    \includegraphics[keepaspectratio, width= 15 cm]{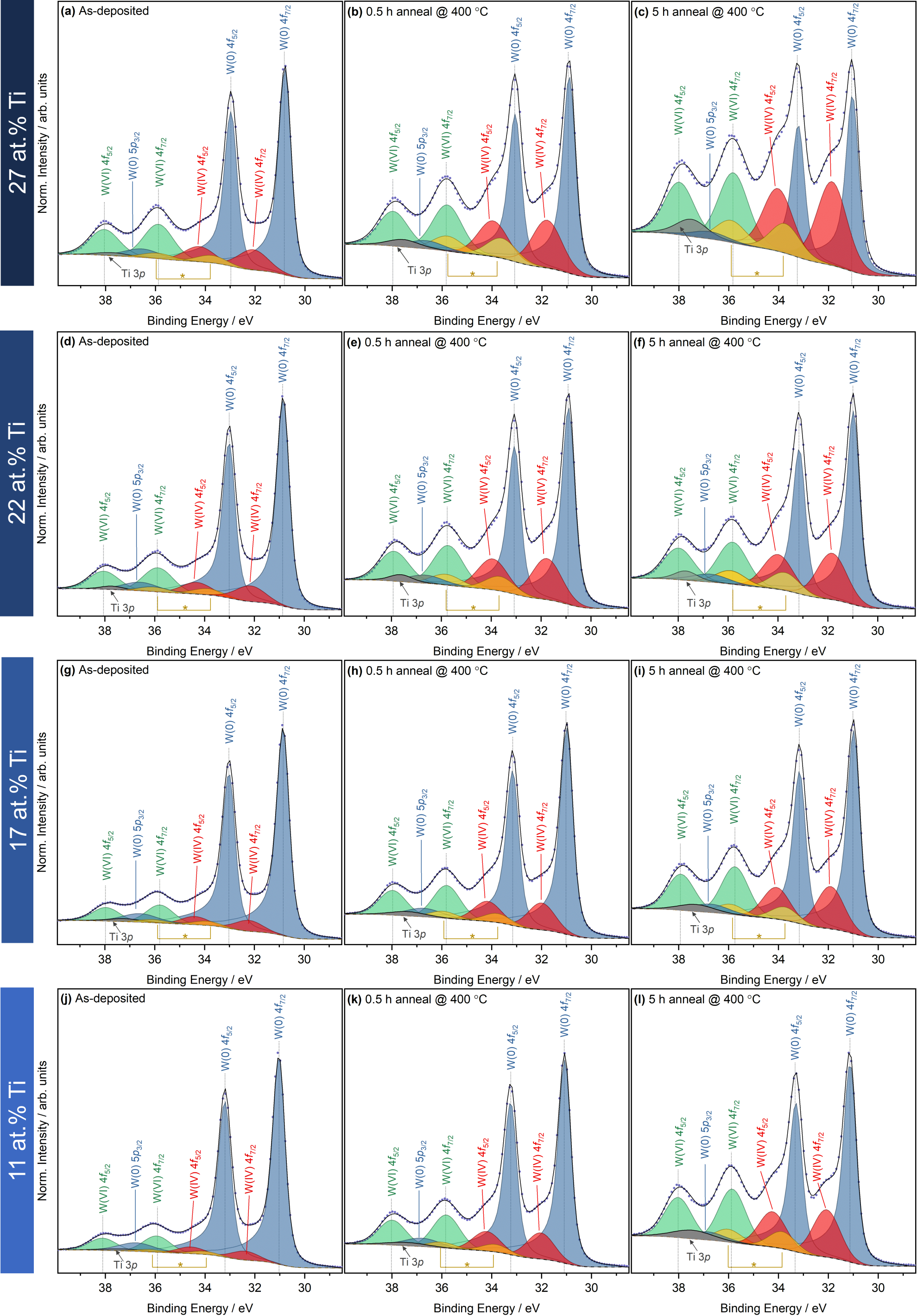}
    \caption{SXPS W~4\textit{f} peak fit summary. Going from left (L) to right (R) - the peaks fits of the as-deposited, 0.5~h anneal and 5~h anneal spectra are presented. Each row displays the spectra for a given titanium composition (11, 17, 22 and 27~at.\%).}
    \label{fig:peakfit}
\end{figure*}

In order to determine the number of oxide contributions in the W~4\textit{f} and Ti~2\textit{p} core levels, as well as quantifying the extent of oxidation, peak fit analysis was conducted on the SXP spectra collected for the as-received (Etch 0) samples. The resultant W~4\textit{f} peak fits are displayed in Fig.~\ref{fig:peakfit} for all sample concentration sets. The W~4\textit{f} spectra are complex and the procedures used to complete the peak fit analysis are described in detail in the Supplementary Information. To summarise, the procedure involved taking the metallic line shape from an etched W~4\textit{f} spectra, which consisted of minimal oxide contributions and transferring that line shape across to the as-received (Etch 0) spectra, while adding in a series of potential oxide contributions. The main difficulty was determining the unknown metallic line shapes due to their asymmetry and the convolution of the region between 35-39~eV, in which W(VI)~4\textit{f} doublet peaks, W(0)~5\textit{p}\textsubscript{3/2} and Ti~3\textit{p} peaks are located. All spectra were fitted with a Shirley-type background and all oxide doublet peaks were fitted with a Lorentzian-Gaussian function, consisting of a Lorentzian contribution of less than 30\%.  As mentioned earlier, two tungsten oxide environments were obviously visible, however, in order to acquire a reasonable fit, an additional oxide environment was added between the already ascribed WO\textsubscript{3} and WO\textsubscript{2} states, coloured in yellow and denoted with an asterisk (*). Several cases have been put forward to suggest that this doublet could be attributed to a W(V) state. Corby \textit{et al}.\ attributed the doublet peaks at approximately 34.3~eV and 36.5~eV to a W(V) environment and correlated the presence of W(V) species to the formation of defective WO\textsubscript{3-x} and the presence of oxygen vacancies, during the synthesis of WO\textsubscript{3} nano-structures.\cite{Corby2018} Xie \textit{et al}.\ took a different approach and reduced WO\textsubscript{3} nanowires into a continuum of oxides, attributing the peak at 34.3~eV to the W(V)~4\textit{f}\textsubscript{7/2} peak.\cite{XIE2012112} Whereas, Alay \textit{et al}.\ investigated similar air-exposed TiW alloys using SXPS and determined the presence of metallic tungsten, WO\textsubscript{3} and WO\textsubscript{x} peaks in the W~4\textit{f} spectrum. These WO\textsubscript{x} sub-oxides were located in a similar position to what we have assigned to WO\textsubscript{2}. However, the peak-fit analysis presented by Alay \textit{et al}.\ appears to show minor discrepancies between the resultant fit envelope and the experimental data, falling in line with our findings. This gives further evidence that this third oxide environment (*) is present and needs to be considered in the peak fitting procedure.~\cite{Alay1991}\par

Given that the oxidation process occurred naturally in the ambient environment and that these samples were not synthesised to form an oxide but rather a bimetallic thin film, the formation of W(V) at room-temperature appears unlikely. Additionally, the assignment of this peak to a tungsten oxide hydrate or tungsten hydroxide appears infeasible as these compounds have only been reported to occur upon full submersion of tungsten metal in water.\cite{WARREN1996345} Besides, any surface hydroxide that may form will be in a rather small concentration relative to the oxide environments and so should not generate such a large signal intensity. Therefore, this oxide environment is assumed to be a result of a WO\textsubscript{3} environment in co-ordination with a titanium environment, causing a negative shift in BE position. Due to the polycrystalline and random nature of the TiW alloy, titanium diffusion in response to annealing may not result in uniform distribution across the surface and therefore it is likely that titanium rich regions exist and surround tungsten regions. Moving from left to right in Fig.~\ref{fig:peakfit} displays an increase in annealing duration, whereas going from the top to the bottom displays a decrease in the bulk titanium concentration. It can be observed that the sample with the greatest oxidation is the one consisting of the highest nominal bulk titanium concentration (27~at.\%), coupled with the longest annealing duration (5~h), displayed in Fig.~\ref{fig:peakfit}(c). Reflecting the initial observations, moving to lower bulk titanium concentrations or to shorter annealing timescales reduces the intensity of the tungsten oxide peaks.\par

\begin{figure*}[t!]
\centering
    \includegraphics[keepaspectratio, width= 14 cm]{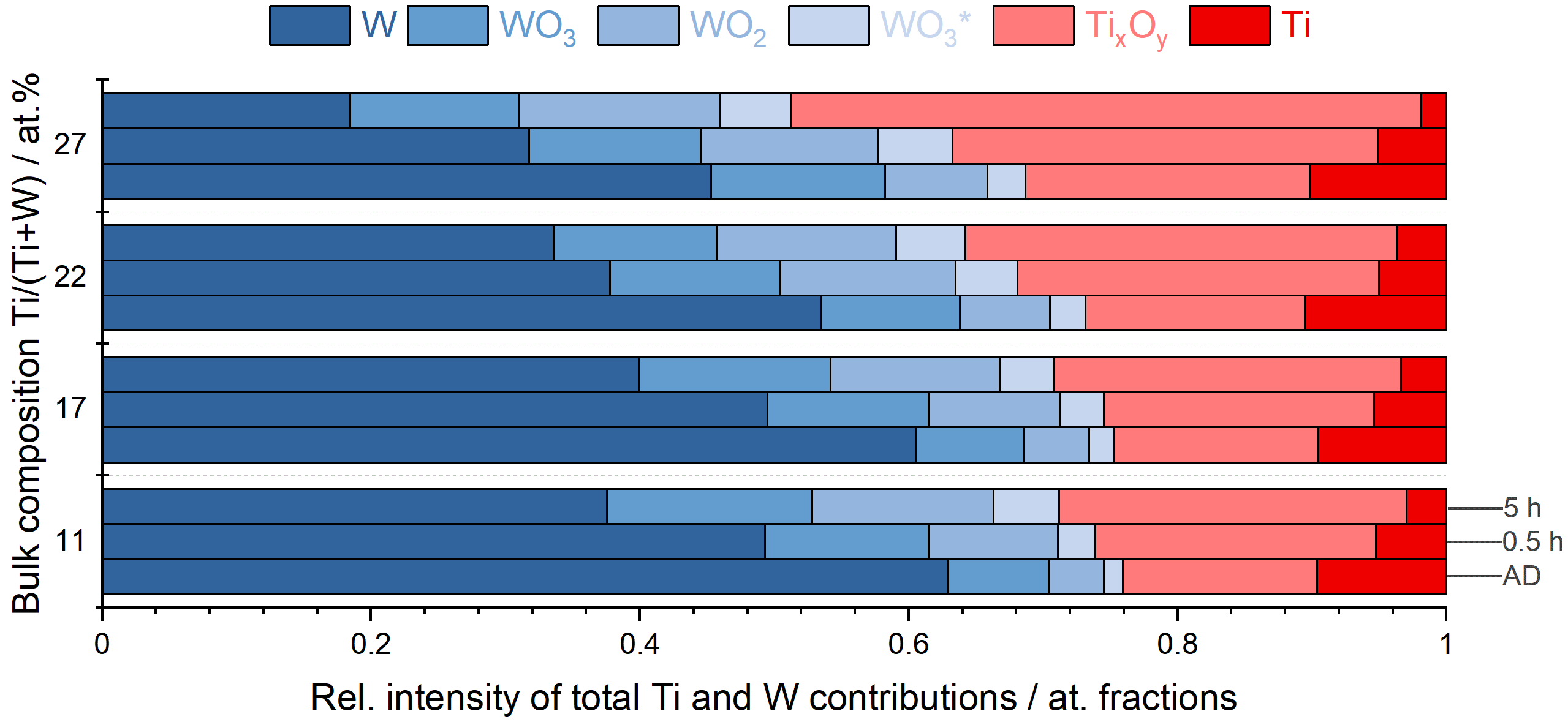}
    \caption{Quantification summary comparing the relative changes between all tungsten and titanium species across all sample sets. The bottom column refers to the AD sample, the middle refers to the 0.5~h annealed sample and the top column refers to the 5~h annealed sampled. WO\textsubscript{3}\textsuperscript{*} refers to the third tungsten oxide environment in co-ordination with a titanium environment. The quantification was obtained from peak fitting the Ti~2\textit{p} and W~4\textit{f} SXP spectra (Etch 0), and by applying the appropriate relative atomic sensitivity factors, photoionisation cross sections, and correction factors. The percentages determined from SXPS peak fit analysis appear similar but are not identical to the listed bulk concentrations. SXPS is extremely surface-sensitive and so will not reflect the bulk composition and additionally, preferential sputtering of the two metal elements may occur, leading to differences between the surface and bulk composition.}
    \label{fig:Quant}
\end{figure*}

Peak fit analysis was also carried out on the Ti~2\textit{p} spectra, with details of the chosen peak model and fitting procedure described in the Supplementary Information. By comparing the peak fitting results of the respective Ti 2\textit{p} and W~4\textit{f} spectra of each sample, quantification between the two metal systems could be established, with the results summarised in Fig.~\ref{fig:Quant}. Due to the complexity of the line shapes of both Ti~2\textit{p} and W~4\textit{f} core lines, it should be noted that the accuracy of quantification from the peak fit analysis is estimated to be approximately $\pm$0.5~at.\%. The comparison revealed that in all cases, tungsten was the dominating component, as expected based on the bulk composition always being tungsten rich. Additionally, the quantification summary visually confirms that with increasing annealing duration comes a decrease in the metallic nature of the TiW surface and the conversion to a more oxide rich system. The rate of oxidation appeared much higher in the case of titanium than tungsten. For the 27~at.\% Ti sample, annealing for 0.5~h resulted in a 37.9\% reduction in the metallic tungsten signal, whereas the metallic titanium signal reduced by 57.3\% relative to the AD sample. The state of the sample surface in the 5~h annealed sample exhibited almost complete transformation from metallic to oxidised titanium, with only 11.8\% of the original metallic signal remaining. On the other hand, the tungsten oxide contribution became larger than the metallic contribution, rising from 34.0\% in the as-deposited sample, to 49.7\% in the 0.5~h annealed sample and, finally surpassing the 50\% mark with a total contribution to the tungsten signal of 64.0\% in the 5~h annealed sample.\par

The idea of competitive oxidation is further reinforced by comparing the WO\textsubscript{3}:WO\textsubscript{2} ratio in Fig.~\ref{fig:Quant}. The as-deposited samples all show that the dominant tungsten oxide contribution is from the WO\textsubscript{3} state. Whereas, for the two annealed samples, the WO\textsubscript{3} concentration appears to hold fairly constant, as if a limit has been reached, while the WO\textsubscript{2} contributions appear to rise. The main assumption regarding the W~4\textit{f} peak model is the assignment of the third tungsten oxide environment. However, the hypothesis that this third environment in which a tungsten oxide environment is in co-ordination with a titanium rich environment holds true, as the signal intensity of these peaks also increases with increasing annealing (i.e. increase in titanium accumulation). In terms of the titanium diffusion process, peak fit analysis revealed that for the sample containing 27~at.\% Ti, an increase of up to $\approx$17~at.\% in the titanium signal was observed after only 0.5~h of annealing, highlighting the severity and speed of this diffusion mechanism. Additionally, by comparing the changes in oxide concentration between the 0.5~h and 5~h annealed samples, it can be observed that the formation of oxides does not appear to follow a linear relationship with annealing duration but instead a parabolic relationship. Metals and alloys often follow a parabolic oxidation rate as the diffusion of oxygen anions and metal cations across the increasing oxide film begins to control the oxide growth rate.\cite{Wagner} Overall the peak fit analysis and results presented suggest that these systems contain many different metal oxide phases, most of which are constrained to the surface, with the bulk appearing more metallic, matching closely with the schematic shown in Fig.~\ref{fig:Diagram}(b).\par

The oxidation of these TiW systems appears to be more impactful towards titanium than tungsten, with almost a complete conversion of metallic titanium to TiO\textsubscript{2} observed in the higher titanium concentration and longer annealed samples. As mentioned throughout the discussion, preferential oxidation toward titanium over tungsten occurs. One possible reason for this mechanism as put forward by Siol \textit{et al}.\ is that the room temperature enthalpy of formation of TiO\textsubscript{2} compared to both WO\textsubscript{3} and WO\textsubscript{2} is lower (TiO\textsubscript{2} = -940.5~kJ/mol,\cite{Kornilov1973} WO\textsubscript{3} = -842.1~kJ/mol\cite{CHARLU1973325}, WO\textsubscript{2} = -586.55~kJ/mol \cite{CHARLU1973325}) which would imply that the generation of TiO\textsubscript{2} would be preferred from the onset and the first to form in the developing oxide layer before WO\textsubscript{3} and WO\textsubscript{2}.\cite{SIOL202095} Additionally, titanium, unlike other transition metals, exhibits a higher solubility of oxygen, with the alpha modification holding up to 32~at.\%.\cite{Kornilov1973} On the other hand, oxygen solubilities for tungsten have been reported as “extremely small” or “very low” and hence do not offer the same oxidation behaviour as titanium.\cite{Wriedt1989} These two factors confirm what is observed in the collected spectra, in that the system exhibits strong preferential oxidation of titanium but the dominance of tungsten in the alloys prevents a significant titanium oxide surface layer to form.\par
Once these samples are exposed to the ambient environment, the oxidation process will immediately occur through physisorption and then chemisorption of oxygen into the alloy. In a pure tungsten system the likelihood of internal oxidation into the bulk is slim, owing to its low oxygen solubility. However, oxidation of tungsten is still thermodynamically feasible according to the reported enthalpy of formation energies and so an external oxide scale will likely form. The addition of titanium will enhance the oxidation process due to its higher oxygen solubility. Titanium will increase the partial pressure and concentration of oxygen at the metal/air interface and at the developing metal/metal oxide interface. This will encourage oxidation of the surrounding tungsten, and this assumption is clearly represented in the recorded spectra. Both samples with a greater bulk titanium concentration and samples annealed for a longer duration exhibit a greater presence of tungsten oxide phases. Overall, these results show that a TiW film exposed to the ambient environment forms a mix of titanium and tungsten oxides at the surface, while the bulk is predominately metallic. Additionally, based on the observations drawn from the W~4\textit{f} and Ti~2\textit{p} spectra, it is apparent that the concentration of surface titanium will control the oxidation kinetics.~\cite{Kuznetsov_1992, SIOL202095,KOFSTAD1967449} \par

\subsection{\label{sec:Elec}Electronic Structure}
%%%%%%%%%%%%%%%%%%%%%%%%%%%%%%%%%%%%%%%%%%%%%%%%%%%%%%%%%%%%%%%%%%%%%%%%%%%

\begin{figure}[ht!]
\centering
    \includegraphics[keepaspectratio, width= 8.5 cm]{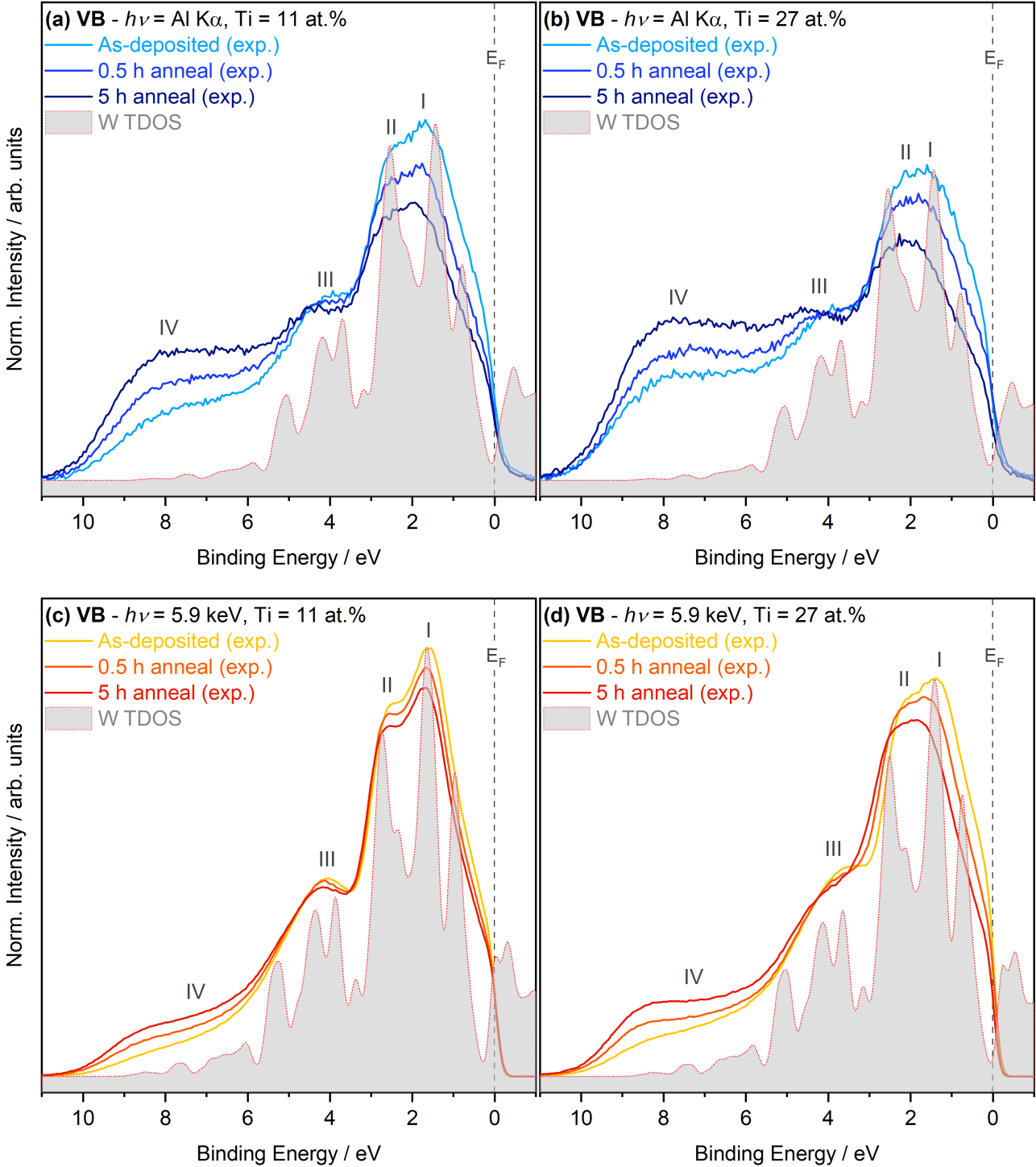}
    \caption{Comparison between the total density of states derived from DFT calculations of W metal with SXPS and HAXPES valence band spectra of TiW samples, including SXPS valence band spectra of (a) 11~at.\% Ti and (b) 27~at.\% Ti, and HAXPES valence band spectra of (c) 11~at.\% and (d) 27~at.\% Ti samples compared to the TDOS of metallic W (W TDOS). Valence band spectra are normalised to their respective areas with a Shirley-type background removed to aid comparison to theory.}
    \label{fig:VB_DFT}
\end{figure}

In parallel to the detailed investigation of the core states, the electronic structure of the occupied valence states of the samples was explored. Fig.~\ref{fig:VB_DFT} displays the valence bands of the 11~at.\% and 27~at.\% Ti samples collected with both SXPS and HAXPES providing surface and bulk information, respectively. Clearly resolved features are denoted with Roman numerals (I-IV) in the figure to aid discussion in the text. Features I, II, III and IV in both spectra are located at approximately 1.4, 2.2, 3.9, and 7.4~eV, respectively. Due to the high concentration of W and the higher photoionisation cross section of the W valence states relative to Ti, the majority contribution to the valence band is attributed to the W 5\textit{d} states.\cite{Scofield1973} Therefore, DFT calculations of metallic W are sufficient, and can be used for comparison with the experimental valence band spectra to aid interpretation. The total and projected density of states, both unweighted and photoionisation cross section weighted for SXPS and HAXPES, can be found in the Supplementary Information. Overall, the comparison of the valence band to the total density of states provided by the DFT calculations are in excellent agreement, particularly for the HAXPES data. It clearly shows that the main contributions to features I-III are due to the W~5\textit{d} states. However, a noticeable disparity between theory and experiment is observed around feature IV at the top of the valence band, particularly in the SXPS data. This can be attributed to the oxidation of the TiW alloy with hybridisation between O~2\textit{p} and W and Ti valence states in full effect. Additionally, with increasing annealing duration and the corresponding increase in Ti concentration at the surface, the intensity in region IV increases, confirming the growing extent of oxidation. At higher annealing duration, a decrease in intensity of the main features I and II is observed, coupled with a reduction in the gradient of the valence band onset at \textit{E}\textsubscript{F}. These effects are a direct response to the increased concentration of titanium following diffusion, subsequently accelerating alloy oxidation behaviour and shifting the metallic states in the valence band towards higher BE.\par

%%%%%%%%%%%%%%%%%%%%%%%%%%%%%%%%%%%%%%%%%%%%%%%%%%%%%%%%%%%%%%%%%%%%%%%%%%%
\section{Conclusions}
%%%%%%%%%%%%%%%%%%%%%%%%%%%%%%%%%%%%%%%%%%%%%%%%%%%%%%%%%%%%%%%%%%%%%%%%%%%

The influence of post-deposition annealing and composition on the oxidation behaviour of TiW thin films was investigated using a combination of SXPS and HAXPES. Surface-sensitive SXPS measurements reveal that longer annealing times result in a greater accumulation of titanium at the TiW/air interface. Consequently, this increases the metal to oxide conversion for both metals. Full metal to oxide conversion was not observed within the SXPS probing depth in either case of titanium or tungsten as metallic features were still retained, suggesting that any oxide is confined predominantly to the surface. Both TiO\textsubscript{2} and Ti\textsubscript{2}O\textsubscript{3} oxides are present, although contributions from the tetravalent oxide dominate. It is also apparent that both WO\textsubscript{2} and WO\textsubscript{3} states exist and increase with annealing duration. However, in order to establish a suitable peak model to allow for accurate quantification a tertiary oxide environment was required and was attributed to peaks between the known WO\textsubscript{2} and WO\textsubscript{3} peaks. This tertiary oxide was assigned to a WO\textsubscript{3} environment in co-ordination with a titanium environment, rather than an intermediate W(V) state. SXPS depth profiling proves useful to probe the elemental composition across the TiW layer, in particular, highlighting the depletion of titanium in the bulk, but the destructive nature of sputtering reduces the ability to acquire detailed information on the bulk oxidation behaviour. However, HAXPES measurements were able to reveal that the bulk of the TiW does experience oxidation and the level of oxidation does increase with increasing annealing duration, due to the migration of titanium. This combinatorial characterisation approach is highly complementary and HAXPES in particular provides a unique insight into the bulk of the TiW film, which is otherwise compromised by the sputtering necessary in conventional SXPS depth profiling. These findings provide sound guidance for the future development of transition metal-based diffusion barriers in semiconductor devices. Furthermore, the results presented here showcase the applicability of X-ray photoelectron spectroscopy approaches to the characterisation of multi-metal, layered systems in general.\par

\section*{Supplementary Material}
See Supplementary Material for SXPS and HAXPES survey spectra, details on peak fitting strategy and results, C~1\textit{s} core level spectra, and the unweighted and cross section weighted projected densities of states from DFT. 

\begin{acknowledgments}
CK acknowledges the support from the Department of Chemistry, UCL. NKF  acknowledges  support  from  the Engineering  and  Physical  Sciences Research  Council  (EP/L015277/1). JJGM and SM acknowledge the support from the FusionCAT project (001-P-001722) co-financed by the European Union Regional Development Fund within the framework of the ERDF Operational Program of Catalonia 2014-2020 with a grant of 50\% of total cost eligible, the access to computational resources at MareNostrum and the technical support provided by BSC (RES-QS-2020-3-0026). LER acknowledges support from an EPSRC Early Career Research Fellowship (EP/P033253/1) and the Thomas Young Centre under grant number TYC-101. AR acknowledges the support from the Analytical Chemistry Trust Fund for her CAMS-UK Fellowship. We acknowledge Diamond Light Source for time on Beamline I09 under Proposal SI19885-1. The authors would like to thank Dave McCue, I09 beamline technician, for his support of the experiments.
\end{acknowledgments}

\section*{Data availability statement}
The majority of the data that supports the findings of this study are available within the article and its Supplementary Material. All data concerning the DFT calculations can be found here: https://b2drop.eudat.eu/s/NrYfWNjbriYJpPm. Any further supporting data are available from the corresponding author.

\section*{Disclosures}
The authors declare no conflicts of interest.

\section{References}

\bibliography{References}
\bibliographystyle{apsrev4-1.bst}

\end{document}

%\preprint{AIP/123-QED}

\title{Thermal and Oxidation Stability of Ti\textsubscript{x}W\textsubscript{1-x} Diffusion Barriers Investigated by Soft and Hard X-ray Photoelectron Spectroscopy\\ Supplementary Information}
% Force line breaks with \\

\author{C. Kalha}
\affiliation{Department of Chemistry, University College London, 20 Gordon Street, London, WC1H~0AJ, United Kingdom.}

\author{S. Bichelmaier}%
\altaffiliation[Current address: ]{Technische Universität Wien, Department of Computational Chemistry, Getreidemarkt 9/165, 1060 Vienna, Austria.}
\affiliation{Kompetenzzentrum Automobil- und Industrie-Elektronik GmbH, Europastraße 8, 9524 Villach, Austria.}

\author{N.~K.~Fernando}
\affiliation{Department of Chemistry, University College London, 20 Gordon Street, London, WC1H~0AJ, United Kingdom.}%Lines break automatically or can be forced with \\

\author{J.~V.~Berens}
\affiliation{Kompetenzzentrum Automobil- und Industrie-Elektronik GmbH, Europastraße 8, 9524 Villach, Austria.}

\author{P.~K.~Thakur}%
\author{T.~-L.~Lee}%
\affiliation{Diamond Light Source Ltd., Harwell Science and Innovation Campus, Didcot, Oxfordshire, OX1 3QR, United Kingdom.}

\author{J.~J.~Gutiérrez Moreno}
\affiliation{Barcelona Supercomputing Center (BSC).}

\author{S.~Mohr}
\affiliation{Barcelona Supercomputing Center (BSC).}
\affiliation{Nextmol (Bytelab Solutions SL), C/ Roc Boronat 117, 08018 Barcelona, Spain.}

\author{L.~E.~Ratcliff}
\affiliation{Department of Materials, Imperial College London, London SW7 2AZ, United Kingdom.}

\author{M.~Reisinger}
\author{J.~Zechner}
\author{M.~Nelhiebel}
\affiliation{Kompetenzzentrum Automobil- und Industrie-Elektronik GmbH, Europastraße 8, 9524 Villach, Austria.}

\author{A.~Regoutz}
 \email{a.regoutz@ucl.ac.uk}
\affiliation{Department of Chemistry, University College London, 20 Gordon Street, London, WC1H~0AJ, United Kingdom.}

\date{\today}% It is always \today, today,

\maketitle

    %%%%%%%%%%%%%%%%%%%%%%%%%%%%%%%%%%%%%%%%%%%%%%%%%%%%%%%%%%%%%%%%%%%%%%%%%%%
    \section{\label{sec:Survey_Spectra}Survey Spectra}
    %%%%%%%%%%%%%%%%%%%%%%%%%%%%%%%%%%%%%%%%%%%%%%%%%%%%%%%%%%%%%%%%%%%%%%%%%%%

\begin{figure}[H]
\centering
    \includegraphics[keepaspectratio, height = 7 cm]{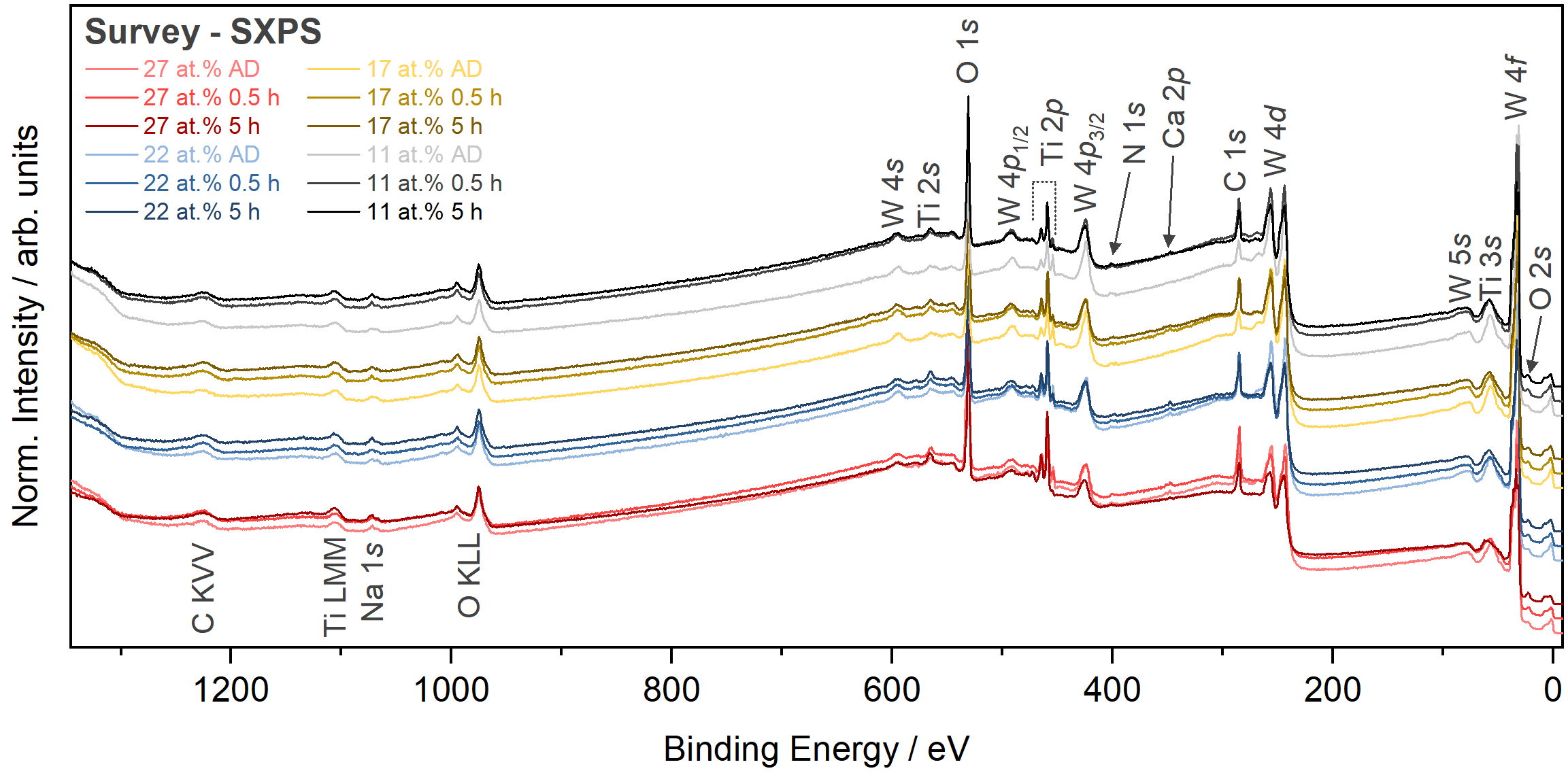}
    \caption{SXPS survey spectra collected for the as-received (i.e.\ no etch) sample sets. Spectra are vertically stacked and normalised to the maximum peak intensity.}
    \label{fig:SI_SXPS_survey}
\end{figure}

\begin{figure}[H]
\centering
    \includegraphics[keepaspectratio, height = 7 cm]{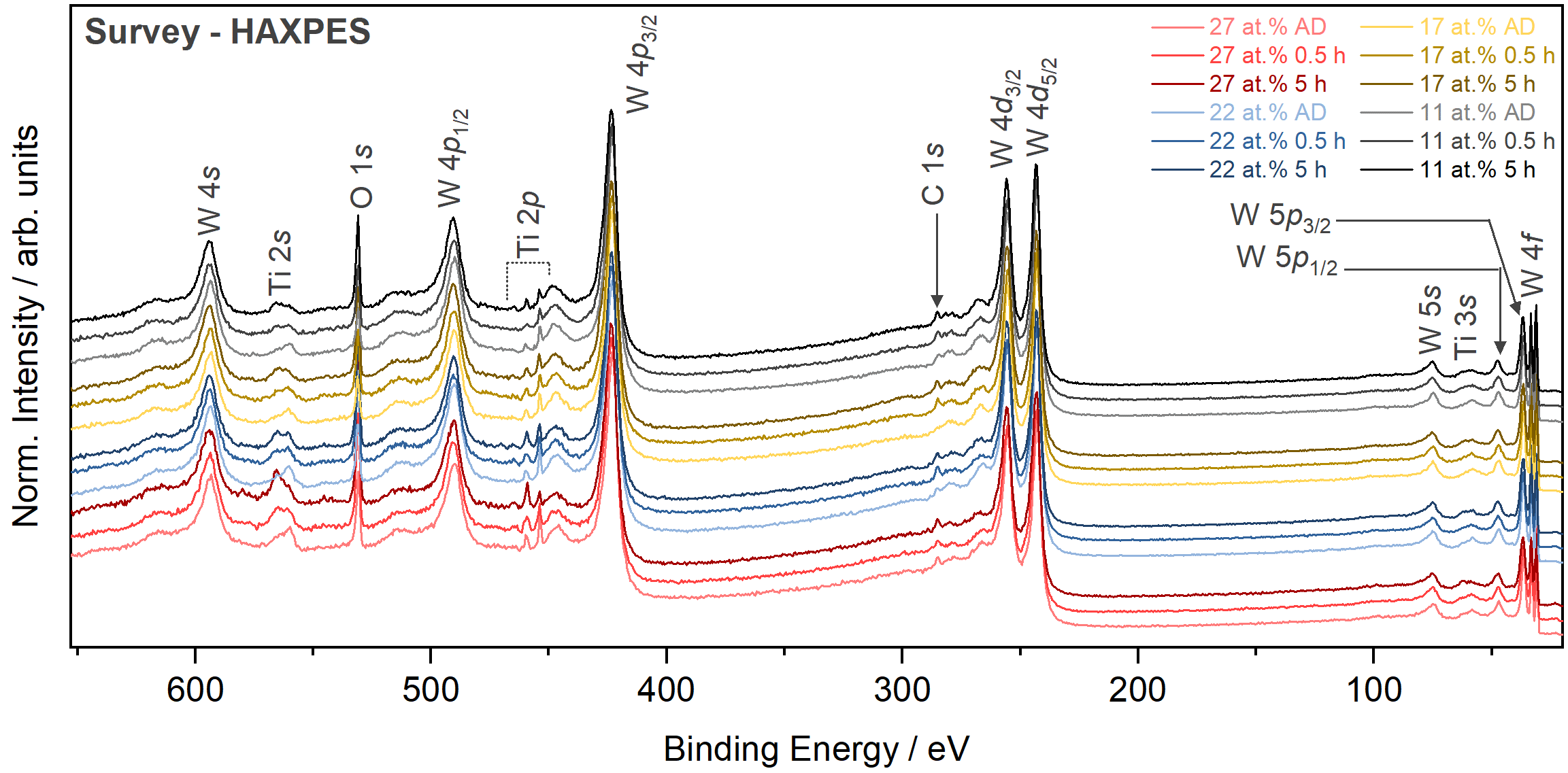}
    \caption{HAXPES survey spectra collected for the as-received sample sets. Spectra are vertically stacked and normalised to the maximum peak intensity.}
    \label{fig:SI_HAXPES_survey}
\end{figure}

Figs.~\ref{fig:SI_SXPS_survey} and~\ref{fig:SI_HAXPES_survey} show the survey spectra collected using SXPS and HAXPES of the as-received samples, respectively. All observed core and Auger lines are indicated.\\

    %%%%%%%%%%%%%%%%%%%%%%%%%%%%%%%%%%%%%%%%%%%%%%%%%%%%%%%%%%%%%%%%%%%%%%%%%%%
    \section{\label{sec:W_4f}SXPS W 4\MakeLowercase{\textit{f}} Binding Energy Positions}
    %%%%%%%%%%%%%%%%%%%%%%%%%%%%%%%%%%%%%%%%%%%%%%%%%%%%%%%%%%%%%%%%%%%%%%%%%%%

The following Tabs.~\ref{SI_27_W4f_BE}-\ref{SI_11_W4f_BE} list the BE positions of all peaks associated with the SXPS W 4\textit{f} core level across all sample sets as determined by peak fit analysis.\\

\begin{table}[H]
\caption{\label{SI_27_W4f_BE}Binding energy (BE) positions of peaks in the  W 4\textit{f} core level region for  samples containing a Ti bulk concentration of 27 at.\%. BE positions determined via peak fit analysis.}
\begin{ruledtabular}
\begin{tabularx}{\textwidth}{cccc}
$Peak$ & $AD~/~eV$ & $0.5~h~anneal~/~eV$ & $5~h~anneal~/~eV$\\
\hline
W(0) 4\textit{f}\textsubscript{7/2} & 30.92 & 31.08 & 31.23 \\
W(0) 4\textit{f}\textsubscript{5/2} & 33.09 & 33.25 & 33.40 \\
W(VI) 4\textit{f}\textsubscript{7/2} & 36.00 & 35.97 & 36.01 \\
W(VI) 4\textit{f}\textsubscript{5/2} & 38.17 & 38.14 & 38.18 \\
W(IV) 4\textit{f}\textsubscript{7/2} & 32.20 & 31.97 & 32.05 \\
W(IV) 4\textit{f}\textsubscript{5/2} & 34.37 & 34.14 & 34.22 \\
W(VI*) 4\textit{f}\textsubscript{7/2} & 33.86 & 33.79 & 33.93 \\
W(VI*) 4\textit{f}\textsubscript{5/2} & 36.03 & 35.96 & 36.10 \\
W(0) 5\textit{p}\textsubscript{3/2} & 36.66 & 36.82 & 36.97 \\
Ti 3\textit{p} (oxide) & 37.85 & 37.73 & 37.69 \\
\end{tabularx}
\end{ruledtabular}
\end{table}

\begin{table}[H]
\caption{\label{SI_22_W4f_BE}Binding energy (BE) positions of peaks in the  W 4\textit{f} core level region for  samples containing a Ti bulk concentration of 22 at.\%. BE positions determined via peak fit analysis.}
\begin{ruledtabular}
\begin{tabular}{cccc}
$Peak$ & $AD~/~eV$ & $0.5~h~anneal~/~eV$ & $5~h~anneal~/~eV$\\
\hline
W(0) 4\textit{f}\textsubscript{7/2} & 30.99 & 31.20 & 31.19 \\
W(0) 4\textit{f}\textsubscript{5/2} & 33.16 & 33.37 & 33.36 \\
W(VI) 4\textit{f}\textsubscript{7/2} & 36.02 & 36.02 & 36.03 \\
W(VI) 4\textit{f}\textsubscript{5/2} & 38.19 & 38.19 & 38.20 \\
W(IV) 4\textit{f}\textsubscript{7/2} & 32.30 & 32.07 & 32.05 \\
W(IV) 4\textit{f}\textsubscript{5/2} & 34.47 & 34.24 & 34.22 \\
W(VI*) 4\textit{f}\textsubscript{7/2} & 34.01 & 33.94 & 33.94 \\
W(VI*) 4\textit{f}\textsubscript{5/2} & 36.18 & 36.11 & 36.11 \\
W(0) 5\textit{p}\textsubscript{3/2} & 36.73 & 36.94 & 36.93 \\
Ti 3\textit{p} (oxide) & 37.86 & 37.93 & 37.91 \\
\end{tabular}
\end{ruledtabular}
\end{table}

\begin{table}[H]
\caption{\label{SI_17_W4f_BE}Binding energy (BE) positions of peaks in the  W 4\textit{f} core level region for  samples containing a Ti bulk concentration of 17 at.\%. BE positions determined via peak fit analysis.}
\begin{ruledtabular}
\begin{tabular}{cccc}
$Peak$ & $AD~/~eV$ & $0.5~h~anneal~/~eV$ & $5~h~anneal~/~eV$\\
\hline
W(0) 4\textit{f}\textsubscript{7/2} & 31.15 & 31.27 & 31.38 \\
W(0) 4\textit{f}\textsubscript{5/2} & 33.32 & 33.44 & 33.55 \\
W(VI) 4\textit{f}\textsubscript{7/2} & 36.09 & 36.08 & 36.14 \\
W(VI) 4\textit{f}\textsubscript{5/2} & 38.26 & 38.25 & 38.31 \\
W(IV) 4\textit{f}\textsubscript{7/2} & 32.52 & 32.28 & 32.31 \\
W(IV) 4\textit{f}\textsubscript{5/2} & 34.69 & 34.45 & 34.48 \\
W(VI*) 4\textit{f}\textsubscript{7/2} & 34.20 & 34.05 & 34.15 \\
W(VI*) 4\textit{f}\textsubscript{5/2} & 36.37 & 36.22 & 36.32 \\
W(0) 5\textit{p}\textsubscript{3/2} & 36.89 & 37.01 & 37.12 \\
Ti 3\textit{p} (oxide) & 37.50 & 37.65 & 37.71 \\
\end{tabular}
\end{ruledtabular}
\end{table}

\begin{table}[H]
\caption{\label{SI_11_W4f_BE}Binding energy (BE) positions of peaks in the  W 4\textit{f} core level region for  samples containing a Ti bulk concentration of 11 at.\%. BE positions determined via peak fit analysis.}
\begin{ruledtabular}
\begin{tabular}{cccc}
$Peak$ & $AD~/~eV$ & $0.5~h~anneal~/~eV$ & $5~h~anneal~/~eV$\\
\hline
W(0) 4\textit{f}\textsubscript{7/2} & 30.95 & 31.15 & 31.23 \\
W(0) 4\textit{f}\textsubscript{5/2} & 33.12 & 33.32 & 33.40 \\
W(VI) 4\textit{f}\textsubscript{7/2} & 35.86 & 35.90 & 35.95 \\
W(VI) 4\textit{f}\textsubscript{5/2} & 38.03 & 38.07 & 38.12 \\
W(IV) 4\textit{f}\textsubscript{7/2} & 32.31 & 32.12 & 32.17 \\
W(IV) 4\textit{f}\textsubscript{5/2} & 34.48 & 34.29 & 34.34 \\
W(VI*) 4\textit{f}\textsubscript{7/2} & 33.96 & 33.92 & 33.95 \\
W(VI*) 4\textit{f}\textsubscript{5/2} & 36.13 & 36.09 & 36.12 \\
W(0) 5\textit{p}\textsubscript{3/2} & 36.69 & 36.89 & 36.97 \\
Ti 3\textit{p} (oxide) & 37.24 & 37.50 & 37.56 \\
\end{tabular}
\end{ruledtabular}
\end{table}

    %%%%%%%%%%%%%%%%%%%%%%%%%%%%%%%%%%%%%%%%%%%%%%%%%%%%%%%%%%%%%%%%%%%%%%%%%%%
    \section{\label{sec:C_1s}C 1\MakeLowercase{\textit{s}} SXPS Core Level Spectra}
\begin{figure}[H]
\centering
    \includegraphics[keepaspectratio, height= 14 cm]{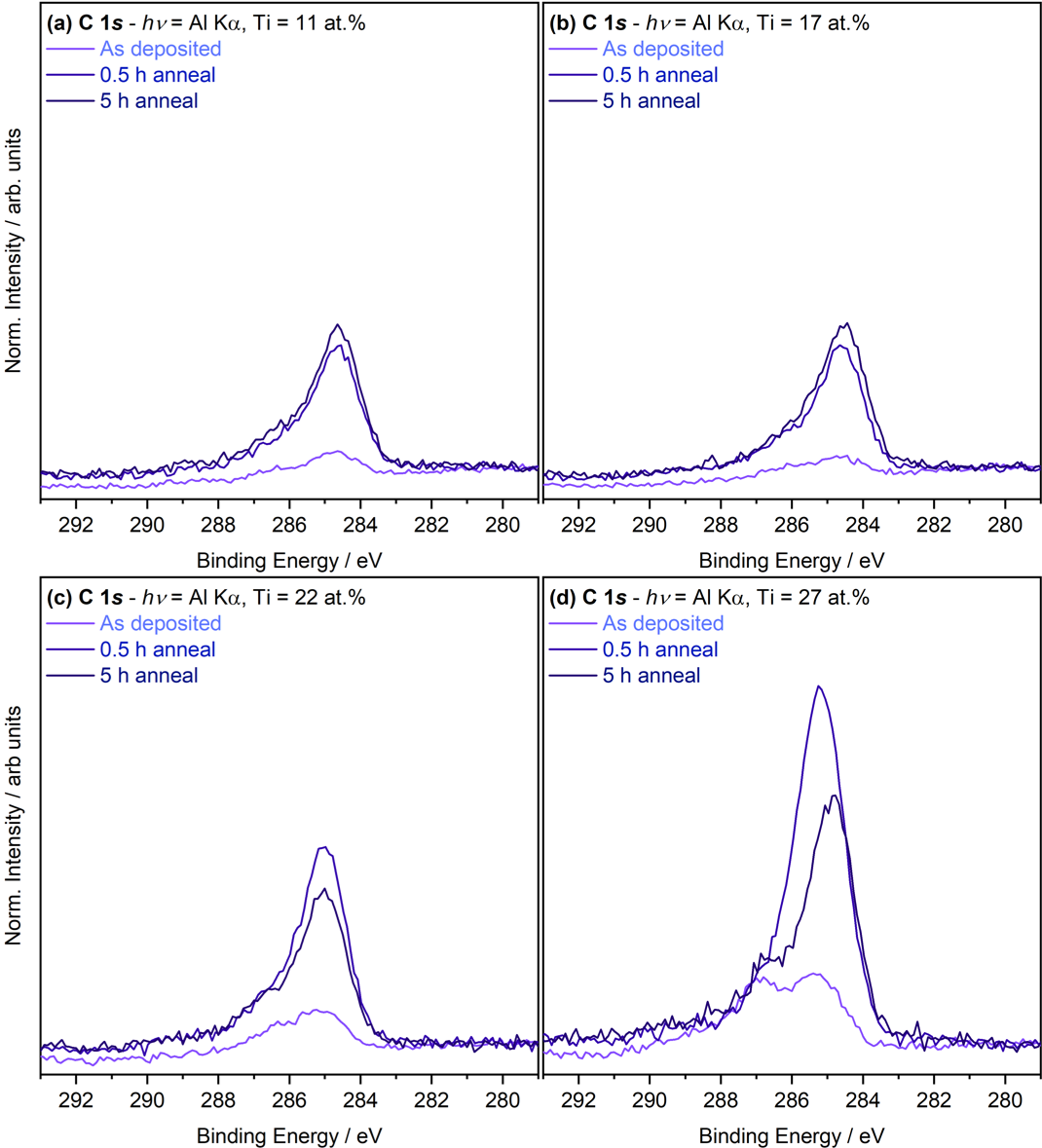}
    \caption{SXPS C 1\textit{s} core level spectra collected for samples containing bulk Ti concentrations of (a) 11 at.\%, (b) 17 at.\%, (c) 22 at.\%, and (d) 27 at.\%, all as a function of annealing duration. All spectra were plotted against the same y-axis range.}
    \label{fig:SI_C1s}
\end{figure}

\begin{figure}[H]
\centering
    \includegraphics[keepaspectratio, height = 7 cm]{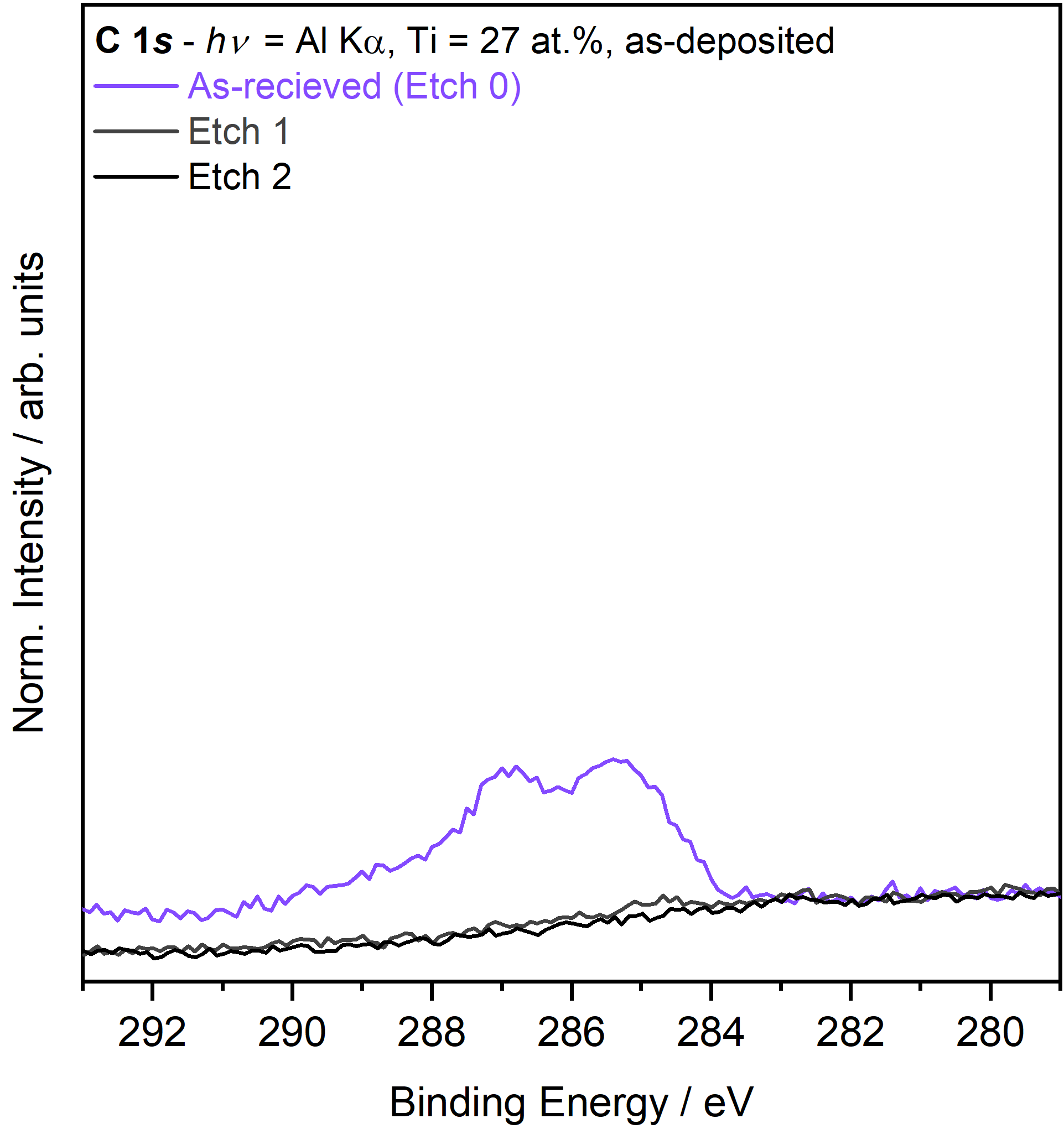}
    \caption{SXPS C 1\textit{s} core level spectra collected for the as-deposited sample containing a bulk titanium concentration of 27 at.\% after each etching step.}
    \label{fig:SI_C1s_etch}
\end{figure}

    %%%%%%%%%%%%%%%%%%%%%%%%%%%%%%%%%%%%%%%%%%%%%%%%%%%%%%%%%%%%%%%%%%%%%%%%%%%
Fig.~\ref{fig:SI_C1s} shows the carbon signals present on the sample surfaces in the SXPS measurements. Multiple C environments are present including both organic (C-C, C-O etc.) and metal carbonate states. In general, samples with a greater bulk Ti concentration possess a greater level of C on the surface. This could be due to the scavenging properties of titanium. Fig.~\ref{fig:SI_C1s_etch} shows that immediately after the first etch step, C is completely removed, indicating that the C detected on as-received samples is a product of surface contamination from the environment, rather than an intrinsic property of the bulk TiW layers. This observation is true for all samples, regardless of the titanium concentration or annealing duration.\\

\FloatBarrier

    %%%%%%%%%%%%%%%%%%%%%%%%%%%%%%%%%%%%%%%%%%%%%%%%%%%%%%%%%%%%%%%%%%%%%%%%%%%
    \section{\label{sec:PF_w}W 4\MakeLowercase{\textit{f}} Peak Fitting Procedure}
    %%%%%%%%%%%%%%%%%%%%%%%%%%%%%%%%%%%%%%%%%%%%%%%%%%%%%%%%%%%%%%%%%%%%%%%%%%%

Peak fit analysis was required to quantify the extent of surface oxidation on the TiW layers. Analysis was conducted on the W 4\textit{f} and Ti 2\textit{p} spectra recorded using SXPS. The laboratory-based SXPS instrument used here can be used to obtain reliable quantification results as it takes into account both the analyser transmission function and relative atomic sensitivity factors (RASFs). Peak fit analysis was applied using the Thermo Scientific\textsuperscript{TM} Avantage software package, which employs very accurate RSFs allowing for the quantification between Ti and W. For all peak fits a Shirley-type background was used and all peaks were fitted with a convolution of Lorentzian and Gaussian functions (L/G mix).\\

\begin{figure*}[h!]
\centering
    \includegraphics[keepaspectratio, width = \linewidth]{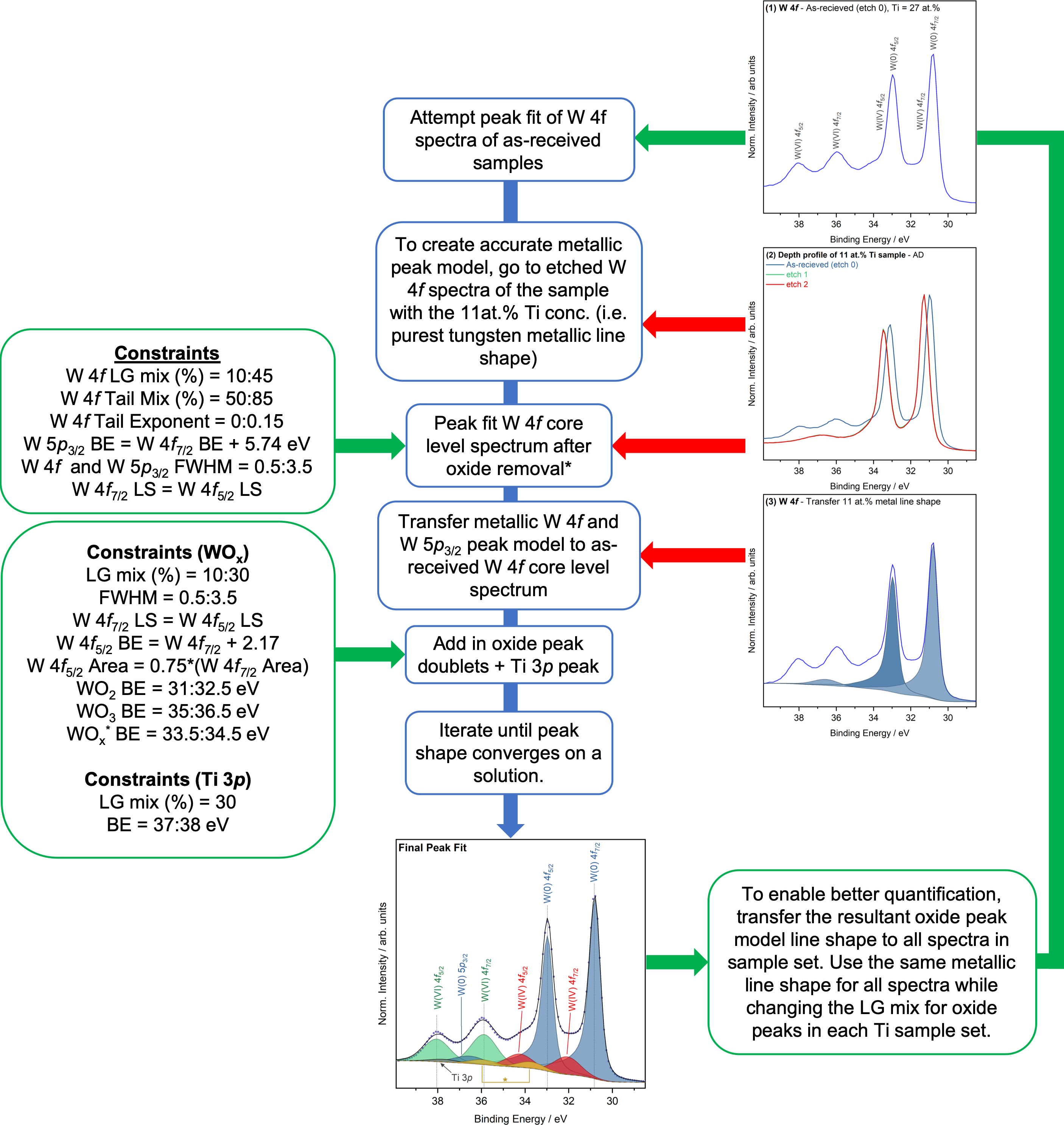}
    \caption{Flowchart process diagram for the SXPS W 4\textit{f} peak fitting procedure, using the fitting of the 27 at.\% Ti as-deposited sample.}
    \label{fig:SI_Flow}
\end{figure*}

Due to the complexity of the W 4\textit{f} spectral line shapes, which arise from the mixture and overlap of both metallic and oxide contributions, the analysis presented many challenges. The following text provides a description of the peak fitting procedure used, with Fig.~\ref{fig:SI_Flow} summarising the process schematically, to enable the application of this procedure to future studies. At least three chemical environments are initially observed in the W 4\textit{f} spectra  - metallic W, WO\textsubscript{2}, and WO\textsubscript{3}. The first major challenge for these peak fits is the mix of lineshapes, with  line-shapes as described by the Kotani model, which are asymmetric with a tail extending toward the higher binding energy side, for metallic states and standard symmetric Voigt profiles for the oxide line-shapes. Additionally, the W 4\textit{f} peaks associated with WO\textsubscript{3} directly overlap the metallic W 5\textit{p}\textsubscript{3/2} peak and the WO\textsubscript{2} W 4\textit{f} peaks sit adjacent to the metallic W 4\textit{f} peaks. Based on the  surface (Etch 0) spectrum alone, the metallic line shape is unknown as it is masked by the oxide contributions, thus preventing the creation of an accurate peak model. However, as these samples were etched in-situ, which involved removing the majority of the oxide contributions, the metallic line-shape can be obtained after etching. By peak fitting the etched W 4\textit{f} spectra belonging to the sample with the greatest tungsten composition, the intrinsic line-shape of the  metallic W contribution could be determined.\\

\begin{figure}[H]
\centering
    \includegraphics[keepaspectratio, height = 7 cm]{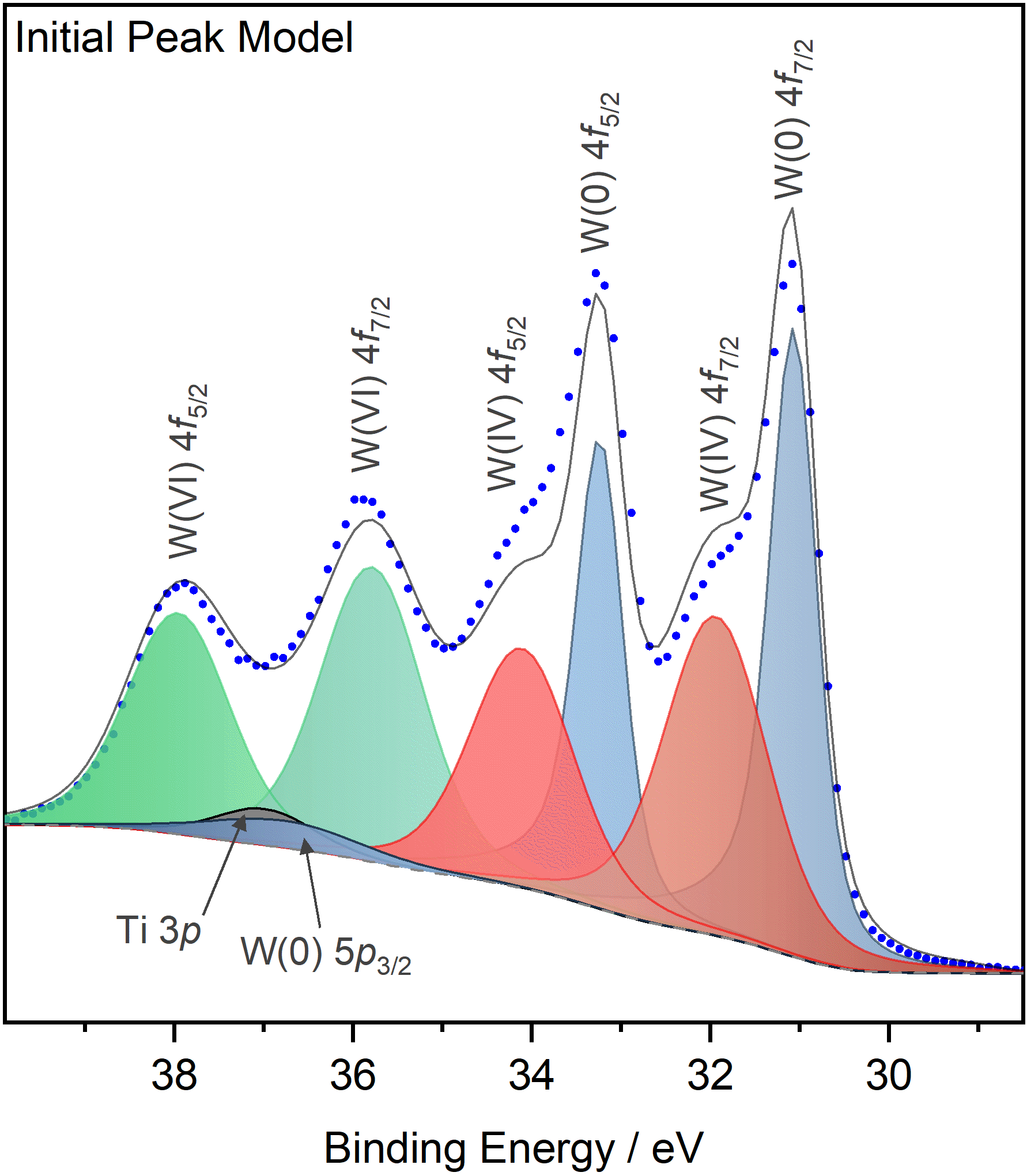}
    \caption{Initial W 4\textit{f} peak fit model with only three environments. A noticeable discrepancy between the fit and the experimental spectrum is observed in the regions directly adjacent to the metallic doublet peaks. This spectra was collected for a 5 h annealed sample with 27 at.\% Ti, measured with SXPS.}
    \label{fig:three-peak}
\end{figure}

The metallic asymmetric tail fitting procedure involved applying an iterative process to determine the tail mix, tail exponent, and L/G mix. Additional constraints were added to ensure that the degeneracy of the doublet W 4\textit{f} peak was maintained, which involved constraining the W 4\textit{f}\textsubscript{5/2} area to be 0.75 times that of the W 4\textit{f}\textsubscript{7/2} area. Additionally, the W 5\textit{p}\textsubscript{3/2} area relative to the metallic W 4\textit{f}\textsubscript{7/2} was constrained, along with the full width at half maximum (FWHM). The LG mix of the W 5\textit{p}\textsubscript{3/2} was set to 30 \% (i.e.\ 70 \% Gaussian, 30 \% Lorentzian). The resultant peak fit metallic model was then transferred to the as-received spectra, as shown in Fig.~\ref{fig:SI_Flow}, ensuring that all the parameters were kept fixed. Once, transferred, the oxide peaks were added. The WO\textsubscript{3} peak positions are known to be approx. 36 eV and 38 eV, positioned on top of the metallic W 5\textit{p}\textsubscript{3/2} peak. The WO\textsubscript{2} peak was assumed to be situated adjacent to the metallic peak. After including the oxide peaks, a noticeable discrepancy between the fit and the experimental spectrum is observed, as shown in Fig.~\ref{fig:three-peak}.\\

To account for the observed mismatch, an additional doublet peak was added. The assumption is that this additional environment corresponds to a WO\textsubscript{3} environment in co-ordination with Ti leading to a negative shift in BE position. The doublet peaks for all oxide contributions were constrained with an LG mix between 10 and 30 \%, additionally, the FWHM of the W 4\textit{f}\textsubscript{5/2} and W 4\textit{f}\textsubscript{7/2} peaks and the FWHM of all oxide environments were constrained to equal each other. The spin orbit splitting between all doublet peaks was fixed at +2.17 eV. To maintain the degeneracy rule, the area of the W 4\textit{f}\textsubscript{5/2} peak was constrained to equal 0.75 times the area of the W 4\textit{f}\textsubscript{7/2} peak. An additional Ti 3\textit{p} peak was also added between the BE range of 37 and 38 eV, as reported values of TiO\textsubscript{2} have been stated within this region. The intensity of this peak is likely to be small given the larger concentration of tungsten and the large cross section of the W 4\textit{f}\textsubscript{7/2} peaks relative to the Ti 3\textit{p} peaks. Additionally, given that the titanium oxide contribution is significantly larger than the metallic contribution in most cases, it was assumed that the metallic Ti 3\textit{p} would not be visible. As the line shape of the Ti 3\textit{p} is unknown, the LG mix was constrained to equal 30 \%.\\

Fig.~\ref{fig:SI_Flow} displays the resultant peak fit. This model gives a reasonable fit to the experimental data. In particular, good agreement is found in the region between 30-35 eV, however, in most cases there is a lack of intensity in the 35-39 eV range. Fitting this region is intrinsically challenging due to the competition between the W(VI) 4\textit{f}\textsubscript{7/2} doublet, Ti 3\textit{p}, and W(0) 5\textit{p}\textsubscript{3/2} peaks. Across the three samples – AD, 0.5 h anneal and 5 h anneal – it was assumed that the metallic line-shape remained constant, whilst the oxide contributions enlarged with increasing annealing. Therefore, the final peak model obtained for the 5 h annealed sample was applied to the other two samples to enable better quantification accuracy across the same set. The peak models used for each sample set (11, 17, 22 and 27 at.\% Ti) are listed in Tab.~\ref{Peak Model} and Tab.~\ref{Oxide Peak Model}.

\begin{table}[H]
\caption{\label{Peak Model}Details of the metallic line shape used for all W 4\textit{f} peak fits acquired from the two-step etched W 4\textit{f} spectra of the as-deposited 11 at.\% Ti sample.}
\begin{ruledtabular}
\begin{tabular}{cccccc}
$\textbf{Peak}$ & $\textbf{FWHM~/~eV}$ & $\textbf{LG~mix~/~\%}$ & $\textbf{Tail~Mix~/~\%}$ & $\textbf{Tail~exponent}$ & $\textbf{Area~Constraint}$\\
\hline
W 4\textit{f} & 0.59 & 42.03 & 82.13 & 0.0755 & W 4\textit{f}\textsubscript{5/2} = W 4\textit{f}\textsubscript{7/2}*0.75\\
W 5\textit{p}\textsubscript{3/2} & 1.70 & 30 & - & - & W 5\textit{p}\textsubscript{3/2} = W 4\textit{f}\textsubscript{7/2}*0.09 \\

\end{tabular}
\end{ruledtabular}
\end{table}

\begin{table}[H]
\caption{\label{Oxide Peak Model}Details of the oxide line shapes used for each sample.}
\begin{ruledtabular}
\begin{tabular}{cccc}
$\textbf{Sample~Ti.~Conc.~/~at.\%}$ & $\textbf{W~4\textit{f}~FWHM~/~eV}$ & $\textbf{W~4\textit{f}~LG~mix~/~\%}$ & $\textbf{Ti~3\textit{p}~FWHM~/~eV}$\\
\hline
27 & 1.41 & 19.26 & 1.46 \\
22 & 1.41 & 12.29 & 1.09 \\
17 & 1.28 & 14.33 & 1.66 \\
11 & 1.29 & 12.53 & 1.86 \\

\end{tabular}
\end{ruledtabular}
\end{table}

The described procedure is deemed the most reliable even if the peak model is still subject to some degree of error, which mainly resides in the accuracy of the metallic line shape model. Even after etching twice, the O 1\textit{s} spectra reveals that lattice oxides are present and therefore Ar sputtering is likely to have reduced the WO\textsubscript{3} and WO\textsubscript{2} states into sub-oxides. Therefore, the assumption that the etched W 4\textit{f} spectra is mostly metallic is subject to some doubt, but given how little oxide remains, the assumption stands as the most plausible hypothesis. As an alternative to the analysis presented here, the less studied W 4\textit{d} core level could be explored, which does not overlap with any Ti core levels, but has a higher lifetime width compared to the W 4\textit{f} states.\\

    %%%%%%%%%%%%%%%%%%%%%%%%%%%%%%%%%%%%%%%%%%%%%%%%%%%%%%%%%%%%%%%%%%%%%%%%%%%
    \section{\label{sec:PF_ti}Ti 2\MakeLowercase{\textit{p}} Peak Fitting Procedure}
    %%%%%%%%%%%%%%%%%%%%%%%%%%%%%%%%%%%%%%%%%%%%%%%%%%%%%%%%%%%%%%%%%%%%%%%%%%%

\begin{figure}[H]
\centering
    \includegraphics[keepaspectratio, height = 7 cm]{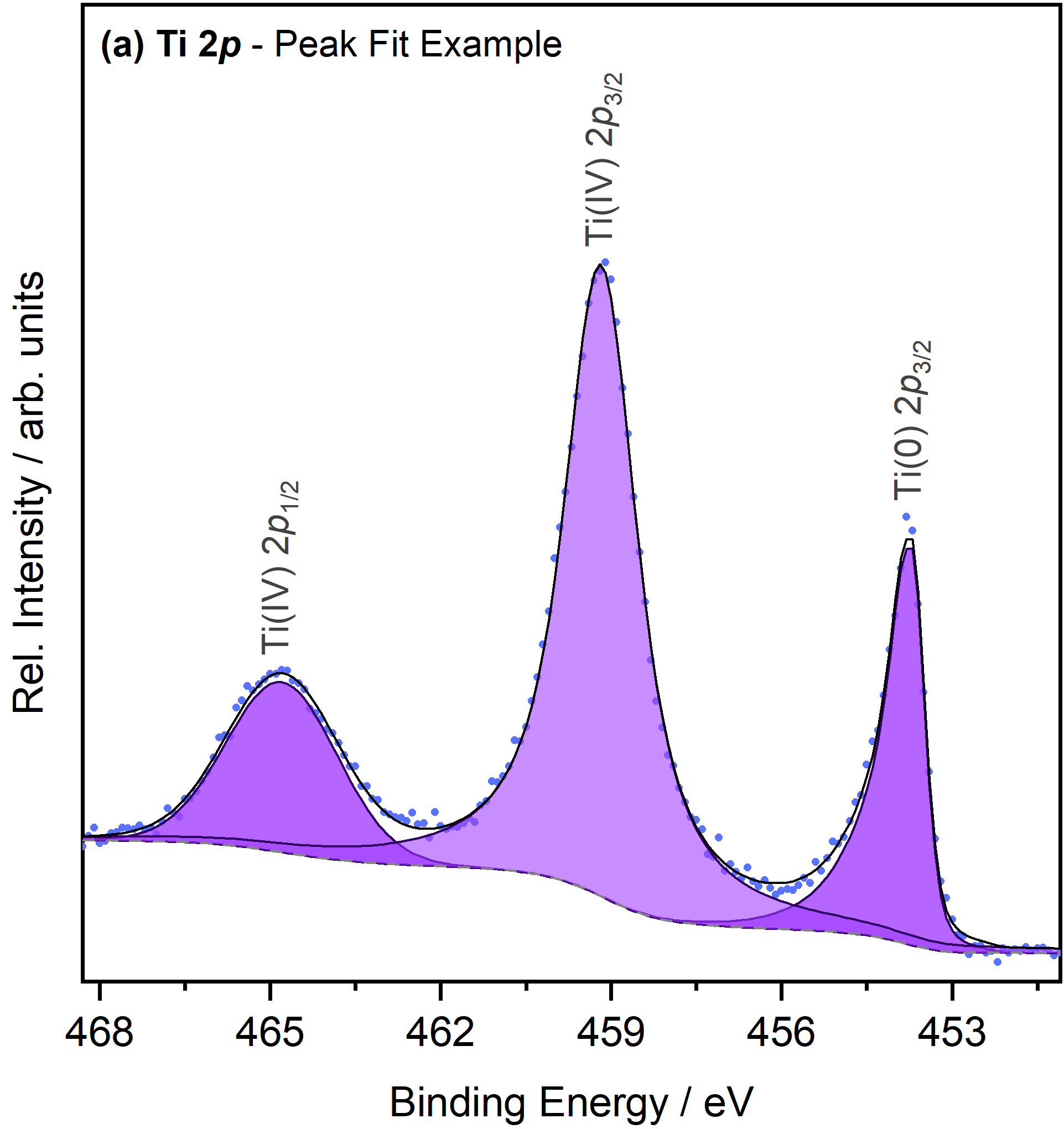}
    \caption{An example peak fit model for the Ti 2\textit{p} core level for the 27 at\% Ti sample (as-deposited) measured with SXPS.}
    \label{fig:Ti}
\end{figure}

Fig.~\ref{fig:Ti} displays SXPS Ti 2\textit{p} core level collected on the higher titanium concentration sample. Peak fitting of the Ti 2\textit{p} spectra is required to determine the extent of surface oxidation. The fitting procedure presents many challenges, most of which are associated with the asymmetric nature of the metallic peaks, the nature of the main intensity Ti 2\textit{p}\textsubscript{3/2} peak, and/or the satellite features hidden underneath the Ti 2\textit{p}\textsubscript{1/2} peak. The metallic Ti 2\textit{p}\textsubscript{1/2} peak is located underneath the Ti(IV) 2\textit{p}\textsubscript{3/2} peak at approximately 460 eV. Due to this overlap, peak fitting was difficult due to the competition between the two peaks, resulting in an increase in Lorentzian nature of the symmetric oxide peak. Therefore, instead of comparing the Ti 2\textit{p}\textsubscript{3/2} peaks to determine the oxidation extent, the metallic Ti 2\textit{p}\textsubscript{3/2} was compared to the Ti(IV) 2\textit{p}\textsubscript{1/2}. To simplify the peak fitting procedure, the main intensity Ti(IV) 2\textit{p}\textsubscript{3/2} peak was modelled as a single peak, ignoring the presence of the underlying Ti(0) 2\textit{p}\textsubscript{1/2}. Additionally, the Ti(IV) 2\textit{p}\textsubscript{1/2} peak was modelled independently to the Ti(IV) 2\textit{p}\textsubscript{3/2} peak, ignoring the underlying Ti 2\textit{p}\textsubscript{3/2} satellite features.\\

    %%%%%%%%%%%%%%%%%%%%%%%%%%%%%%%%%%%%%%%%%%%%%%%%%%%%%%%%%%%%%%%%%%%%%%%%%%%
    \section{\label{sec:PDOS}Tungsten DFT PDOS}
    %%%%%%%%%%%%%%%%%%%%%%%%%%%%%%%%%%%%%%%%%%%%%%%%%%%%%%%%%%%%%%%%%%%%%%%%%%%

\begin{figure}[H]
\centering
    \includegraphics[keepaspectratio, width = \linewidth]{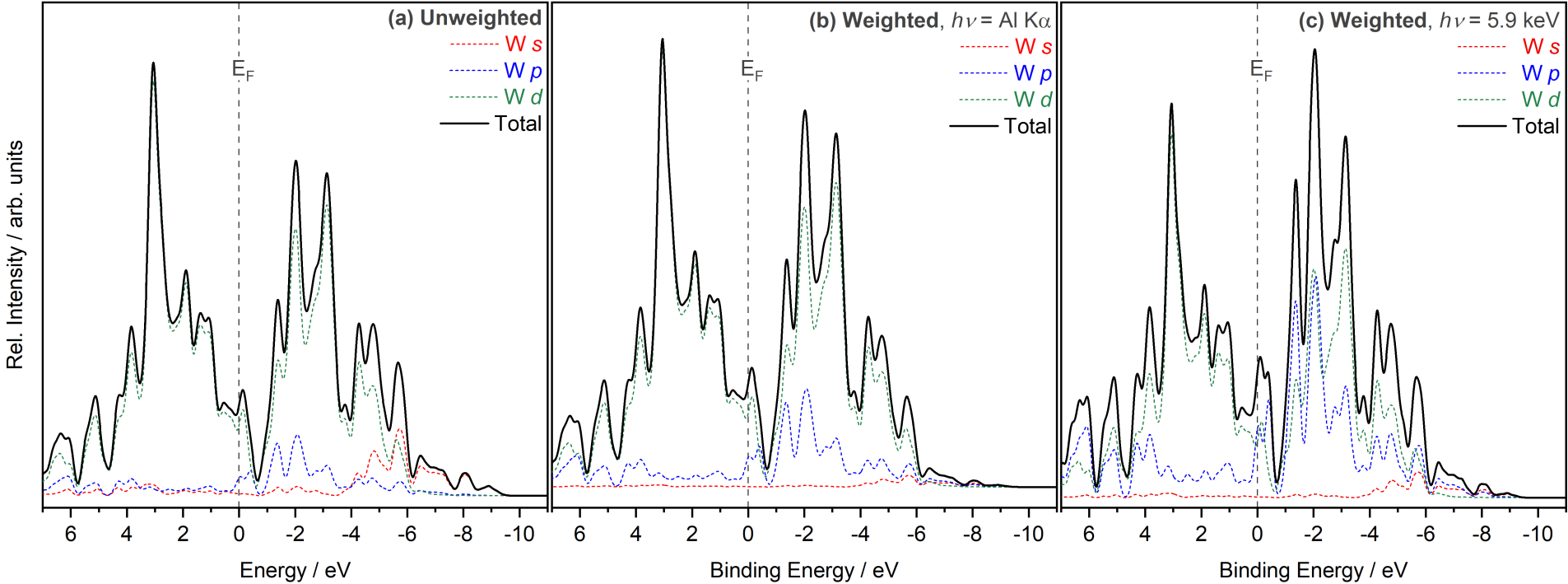}
    \caption{Generated PDOS from DFT calculations. (a) Unweighted PDOS, with a Gaussian smearing of 0.3 ~eV. Broadened and weighted PDOS at photon energies of (b) Al K$\alpha$ and (c) 5.9~keV. The PDOS have been weighted according to the photoionisation cross section values given at the respective experimental photon energies. The PDOS spectra have also been broadened with a Gaussian function to match the broadening from the experimental resolution. }
    \label{fig:PDOS}
\end{figure}

\bibliography{References}
\bibliographystyle{apsrev4-1.bst}